\documentclass[twocolumn]{aastex631}

\usepackage{amsmath}
\usepackage{xspace}
\usepackage{xcolor}
\usepackage{graphicx}
\usepackage{hyperref}
\usepackage{apjfonts}
\usepackage{csvsimple}
\usepackage{longtable}

\definecolor{darkgreen}{rgb}{0.0,0.6,0.0}

\newcommand*{\radius}{inner cluster\xspace}
\newcommand*{\tworadius}{outer cluster\xspace}
\newcommand*{\xone}{\ensuremath{x_1}\xspace}

\begin{document}

\title{Environmental Dependence of Type Ia Supernovae in Low-Redshift Galaxy Clusters}

\correspondingauthor{Conor Larison}
\email{cl1449@physics.rutgers.edu}

\author[0000-0003-2037-4619]{Conor Larison}
\altaffiliation{NSF Graduate Research Fellow}
\affiliation{Department of Physics \& Astronomy, Rutgers, the State University of New Jersey, Piscataway, NJ 08854, USA}

\author[0000-0001-8738-6011]{Saurabh W.~Jha}
\affiliation{Department of Physics \& Astronomy, Rutgers, the State University of New Jersey, Piscataway, NJ 08854, USA}

\author[0000-0003-3108-1328]{Lindsey A.~Kwok}
\affiliation{Department of Physics \& Astronomy, Rutgers, the State University of New Jersey, Piscataway, NJ 08854, USA}

\author[0000-0002-9830-3880]{Yssavo Camacho-Neves}
\affiliation{Department of Physics \& Astronomy, Rutgers, the State University of New Jersey, Piscataway, NJ 08854, USA}

\begin{abstract}

We present an analysis of 102 type Ia supernovae (SNe Ia) in nearby ($z < 0.1$), x-ray selected galaxy clusters. This is the largest such sample to date and is based on archival data primarily from ZTF and ATLAS. We divide our SNe~Ia into an inner cluster sample projected within $r_{500}$ of the cluster center and an outer cluster sample projected between $r_{500}$ and $2\,r_{500}$. We compare these to field samples of SNe~Ia at similar redshifts in both quiescent and star-forming host galaxies. Based on SALT3 fits to the light curves, we find that the inner cluster SNe~Ia have a higher fraction of fast-evolving objects (SALT3 $x_1 < -1$) than the outer cluster or field quiescent samples. This implies an intrinsically different population of SNe~Ia occurs in inner cluster environments, beyond known correlations based on host galaxy alone. Our cluster samples show a strongly bimodal $x_1$ distribution with a fast-evolving component that dominates the inner cluster objects ($\gtrsim$ 75\%) but is just a small fraction of SNe~Ia in field star-forming galaxies ($\lesssim$ 10\%). We do not see strong evidence for variations in the color (SALT3 $c$) distributions among the samples and find only minor differences in SN~Ia standardization parameters and Hubble residuals. We suggest that the age of the stellar population drives the observed distributions, with the oldest populations nearly exclusively producing fast-evolving SNe~Ia.

\end{abstract}


\section{Introduction} \label{sec:intro}

Due to their high and standardizable luminosity \citep{1993ApJ...413L.105P}, type Ia supernovae (SNe Ia) are a key part of the cosmic distance ladder. Measurements of SN~Ia distances led to the discovery of the accelerating expansion of the Universe \citep{1998AJ....116.1009R,1999ApJ...517..565P} and are used to determine the local value of the Hubble constant \citep{2009ApJ...699..539R,2011ApJ...730..119R,2016ApJ...826...56R,2022ApJ...934L...7R,2018ApJ...869...56B,2018A&A...609A..72D,2019ApJ...882...34F}. SNe Ia also contribute to the chemical enrichment of galaxies and are the dominant source of iron-group elements \citep{2013ARA&A..51..457N}. Despite their great importance, the fundamental astrophysics of SNe Ia, including their progenitor channels and explosion mechanisms, is not well understood. Currently, the only consensus is that SNe Ia result from exploding carbon-oxygen white dwarfs \citep[for reviews, see e.g.,][]{Jha:2019,2023RAA....23h2001L}.

The environments of supernovae provide important clues to their astrophysics. For example, the association of core-collapse supernovae with recent star-formation points to a massive star origin. SNe Ia, in contrast, occur in every type of host galaxy, though they are most common in star-forming galaxies. Because star-formation is correlated with other galaxy properties, this also means SNe~Ia occur more frequently in bluer galaxies, morphologically late-type galaxies, and lower-mass galaxies \citep{1990PASP..102.1318V,2005A&A...433..807M,2006ApJ...648..868S,2019MNRAS.484.3785B}. Not only is the SN~Ia rate higher in certain types of host galaxies, the light curve properties of the SNe~Ia are also connected to their environment \citep{1996AJ....112.2391H,2000AJ....120.1479H,1996ApJ...465...73B,2006ApJ...648..868S,2013A&A...560A..66R}. This in turn means that SN~Ia environments are linked to their standardization, and may impact the use of SNe~Ia as cosmological probes, because host galaxy properties vary with redshift. For instance, this could introduce a bias to the dark energy equation of state parameter, \emph{w}, of around its current statistical uncertainty of $\sim$4\% \citep{2020A&A...644A.176R,2021ApJ...909...26B, 2022ApJ...938..110B,2022MNRAS.517.4291D}. The relationship between environment and SN~Ia light curve properties also has important implications for progenitor and explosion models.

Significant evidence has accumulated for an environmental dependence to SN~Ia luminosity, even after light-curve standardization. The first indications of a Hubble residual correlated with host environment were based on the global stellar mass or the star-formation rate of the host galaxy \citep{2010ApJ...715..743K,2010MNRAS.406..782S,2010ApJ...722..566L}. Often a ``mass-step'' is now applied in cosmological analyses to correct for this \citep{2011ApJ...737..102S,2014A&A...568A..22B,2018ApJ...859..101S,2020MNRAS.494.4426S}. Correlations with SN~Ia Hubble residual have also been found using other host-galaxy environmental attributes, including projected separation from the host nucleus, host-galaxy metallicity, and host-galaxy dust content \citep{2011ApJ...743..172D,2012ApJ...755..125G,2023MNRAS.518.1985M}. In addition to ``global'' host properties, SN~Ia luminosity has also been correlated with ``local'' measurements of stellar mass, star-formation rate, and specific star-formation rate \citep[sSFR;][]{2013A&A...560A..66R,2018A&A...615A..68R,2018ApJ...867..108J,2019ApJ...874...32R,2020A&A...644A.176R,2022A&A...657A..22B}. Furthermore, the color of the local SN environment has also been shown to correlate with SN~Ia luminosity \citep{2018A&A...615A..68R, 2021MNRAS.501.4861K, 2023MNRAS.519.3046K,2022A&A...657A..22B}.

Of key importance is understanding the causation behind these correlations. The light curve or luminosity of a SN~Ia is surely not directly influenced by its host-galaxy stellar mass, for example (as noted by \citet{2010MNRAS.406..782S}). Instead, presumably the local or host-galaxy environment is indirectly related to the kinds of white dwarf progenitor systems available to explode. The distributions of metallicity or age of the progenitor population may be the intermediaries that link the environment with the supernova explosion. In fact, it has been noted for a long time that progenitor age may be the main driver of SN~Ia properties \citep{2010MNRAS.406..782S,2010AJ....140..804B,2011ApJ...740...92G}, and further studies have indicated that the age of the stellar population (distinctly correlated with host-galaxy properties described above) may be the dominant factor in shaping the kinds of SNe~Ia that occur \citep{2019ApJ...874...32R,2020ApJ...896L...4R,2020ApJ...903...22L,2020ApJ...889....8K,2021MNRAS.506.3330W,2023MNRAS.520.6214W}.

While the physical causal mechanism may not yet be conclusively known, the empirical correlations between SNe~Ia and their host-galaxy (or smaller) scale environments are well established. It is intriguing to ask then whether these empirical correlations hold at even larger scales. Here we revisit the nature of SNe~Ia found in clusters of galaxies.

Much work has been done on the rate of SNe Ia in cluster host galaxies, which estimates the number of SNe Ia that occur in a galaxy normalized by the galaxy's stellar mass. Early studies of SNe Ia in low-redshift galaxy cluster members show that the SN Ia rate in cluster elliptical hosts is similar to or perhaps elevated compared to the rate in elliptical hosts in the field  \citep{2007ApJ...660.1165S,2008MNRAS.383.1121M,2010ApJ...715.1021D,Maoz:2010,2012ApJ...746..163S}. Higher-redshift galaxy cluster studies have also measured the SN Ia rate in clusters and found similar trends \citep{2010ApJ...718..876S,Barbary:2012,Freundlich:2021,2023MNRAS.526.5292T}. The uncertainties in many of these studies have been dominated by small number statistics, however.

Beyond merely the rate of SNe~Ia in galaxy clusters, it is interesting to compare their light-curve properties and standardized luminosities with SNe~Ia in the field. \citet{2012ApJ...750....1M}, using Hubble Space Telescope (HST) data of high-redshift ($z \simeq 1$) cluster SNe~Ia, found no significant differences with field SNe~Ia, but with only a small sample size of six cluster SNe~Ia with elliptical hosts.
\cite{2013MNRAS.434.1443X} found evidence that SNe~Ia in galaxy clusters (with a sample size of 48 objects) at intermediate redshift ($z \simeq$ 0.1--0.5) had faster-evolving light curves than those in the field, even when restricting both samples to passive galaxies. They ascribed this effect to the older average age of cluster passive galaxies compared to field passive galaxies. Recently, \cite{2023MNRAS.526.5292T} used a larger sample of 70 cluster SNe~Ia at redshifts $z < 0.7$ and also found evidence for faster decline rates compared to field SNe~Ia (though not specifically restricting to only passive galaxies). This trend appears to continue down to low redshift, but published samples are sparse \citep{2004A&A...415..863G}.

Our analysis investigates the properties of nearby ($z < 0.1$) SNe~Ia in x-ray selected galaxy clusters. The x-ray selection assures a higher fidelity cluster sample than the typically optically-selected clusters at higher redshift. Our total sample includes 102 cluster SNe~Ia and hundreds of field objects for comparison, also improving statistics compared to previous work. The advent of large-area time-domain surveys, e.g., PTF \citep{Law:2009}, ASAS-SN \citep{Holoien:2017}, ATLAS \citep{2018PASP..130f4505T}, and ZTF \citep{2019PASP..131a8002B}, allows us to build a large sample of nearby cluster SNe~Ia with multicolor light curves through archival research, rather than requiring a dedicated observing program \citep[e.g.,][]{1998AJ....115...26R,Gal-Yam:2008}.

\section{Data and Methods} \label{sec:data_methods}

\subsection{Galaxy cluster catalog} \label{sec:cluster_catalog}

For this study, we use the Meta Catalog of X-ray Detected Clusters of Galaxies \citep[MCXC;][]{2011A&A...534A.109P} to select a sample of 663 galaxy clusters within our redshift range of interest, $z<0.1$. The catalog also includes information about the cluster sizes, x-ray luminosities, and inferred masses. Specifically, we rely on the catalog $r_{500}$ measurement, the radius of the cluster at which the mass density is 500 times the critical density of the Universe at the cluster redshift. The MCXC $r_{500}$ values depend on assumptions about cluster relations that are detailed by \citet{2011A&A...534A.109P}, and the catalog adopts a flat ${\Lambda}$CDM cosmology with $H_{0} = 70$  km$\,$s$^{-1}\,$Mpc$^{-1}$, $\Omega_{M} = 0.3$, and $\Omega_\Lambda = 0.7$. We adopt this cosmological model in our analysis for consistency.

\subsection{Supernova samples} \label{sec:sample_selections}

\begin{figure}
    \centering
    \includegraphics[scale=0.21]{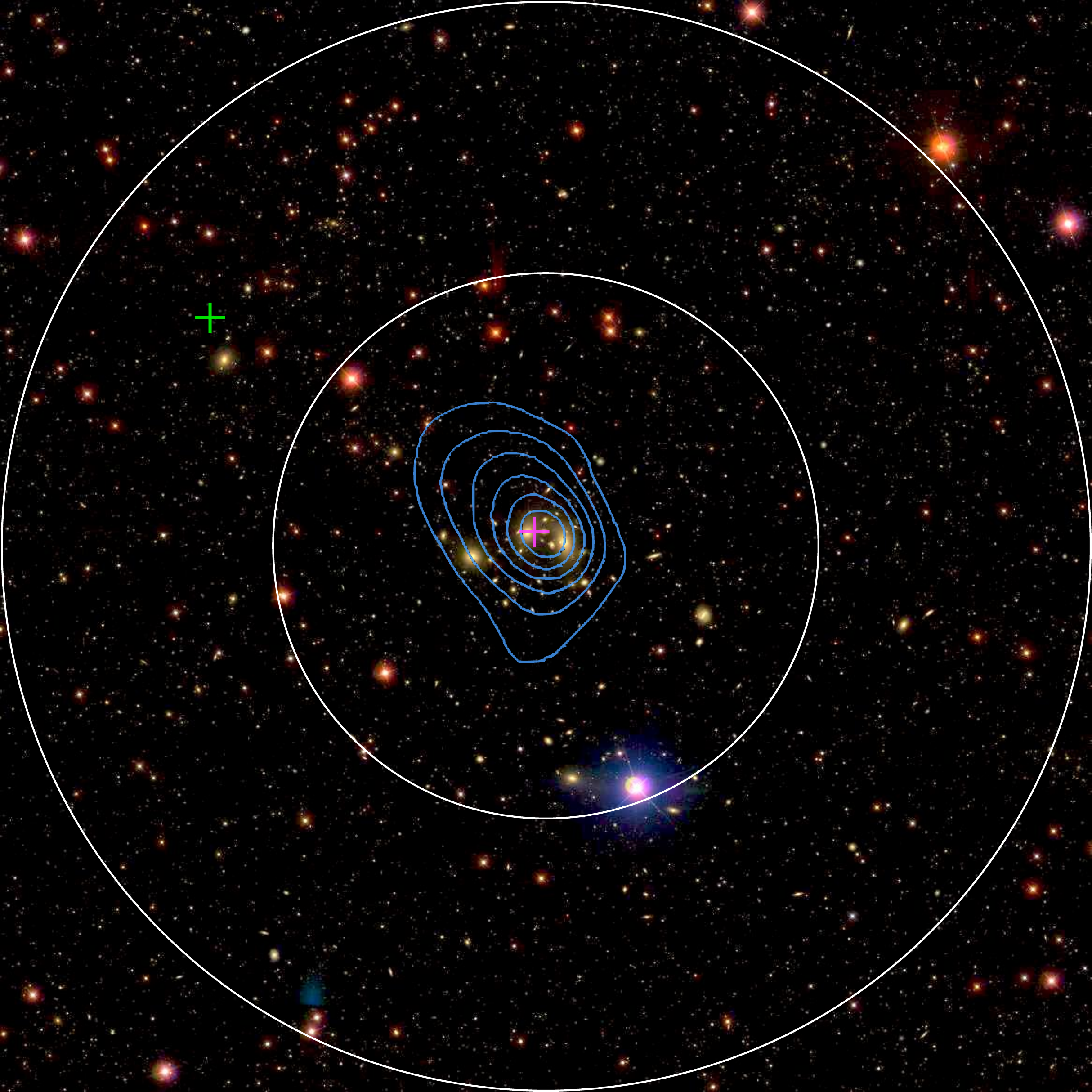}
    \caption{SDSS optical (color) and ROSAT PSPC x-ray (contours) image of the galaxy cluster MCXC J2310.4$+$0734 at $z = 0.0424$. The inner white circle corresponds to $r_{500} =$ 0.73 Mpc (0.24$^\circ$ radius) and the outer circle is twice that radius, roughly the virial radius of the cluster. North is up and east is to the left. The position of the type-Ia SN~2020acwj, part of our \tworadius sample, is shown in green and the position of SN~2021wyw, a fast-declining SN~Ia in our \radius sample, is shown near the center of the cluster in pink.}
    \label{fig:cluster_example}
\end{figure}

To build our SN Ia samples, we select SNe projected within $2r_{500}$ of each cluster, converting the MCXC tabulated $r_{500}$ from a physical to angular size using the angular diameter distance appropriate for the cluster redshift. We split the cluster SN~Ia sample into an \radius\ sample, for SNe~Ia within $r_{500}$, and an \tworadius\ sample between $r_{500}$ and $2r_{500}$. The \radius\ sample probes the centers of our clusters, typically including the extent of observed x-ray emission, and populated by mainly early-type, quiescent galaxies \citep{1985ApJ...292..404G}. The \tworadius\ sample includes SNe~Ia that extend out to approximately the virial radius of the clusters \citep{2013SSRv..177..195R,2019SSRv..215....7W}. An example of one of the clusters in our sample, hosting two SNe~Ia, is shown in \autoref{fig:cluster_example}. We reiterate that our sample division is based on the projected separation; we discuss below how we estimate cluster membership and contamination below and in \autoref{sec:sample_comparison}.

We identify our supernova samples by querying the Transient Name Server\footnote{https://www.wis-tns.org/} (for objects discovered after 2016) and the IAU List of Supernovae\footnote{http://www.cbat.eps.harvard.edu/lists/Supernovae.html} for older objects. We restrict our sample to SNe that have been spectroscopically classified as regular SNe Ia, and we check the classification by manually inspecting light curves (see \autoref{sec:photometric_data}). 

We use the ``directional light radius" \citep[DLR;][]{2006ApJ...648..868S,2016AJ....152..154G} method and the Galaxies HOsting Supernova Transients (GHOST)
database \citep{2021ApJ...908..170G} to associate each potential cluster SN Ia with a host galaxy matched against the NASA Extragalactic Database (NED)\footnote{The NASA/IPAC Extragalactic Database (NED) is operated by the Jet Propulsion Laboratory, California Institute of Technology, under contract with the National Aeronautics and Space Administration.} or SIMBAD \citep{2000A&AS..143....9W}. These account for $\sim$75\% and  67\% of our inner and outer cluster sample host identifications respectively. For the remainder, we lack the requisite imaging data to measure the galaxy photometry and light profiles, but can unambiguously identify the host by visual inspection. In our \radius sample, there were two supernovae, SN~2020wcj and SN~2020yji, that did not have obviously identifiable hosts, while in our \tworadius sample, there were three such supernovae: SN~2020ags, SN~2020vnr, and SN~2022rdt.

Each SN Ia that had an identifiable host was associated with a NED source: either the host galaxy itself, or absent that, a WISE source \citep{Cutri:2021}, from which we collated photometry. We adopted the NED host-galaxy spectroscopic redshift if available (the majority of objects), or else we used the redshift from the SN spectrum as reported in the supernova discovery or classification. Following \cite{2022PASA...39...46C}, we adopt redshift uncertainties of $\sigma_z = 0.0001$ or $\sigma_z = 0.005$ for host-galaxy spectroscopic redshifts or SN spectrum redshifts, respectively. If only a host photometric redshift was available, we adopt $\sigma_z = 0.01$.

We use these redshifts to verify that each SN host galaxy is a member of its cluster. We follow \cite{2013MNRAS.434.1443X} and calculate the membership probability with
\begin{equation}
    p = \frac{1}{\sqrt{2 \pi \left(\sigma_{\text{SN}}^2 + \sigma_{\text{CL}}^2\right)}}
    \int_{-z_d}^{+z_d} \exp 
    \left[ -\frac{\left(z - \left[z_{\text{SN}} - z_{\text{CL}}\right]\right)^2}{2 \left(\sigma_{\text{SN}}^2 + \sigma_{\text{CL}}^2\right)}\right] \, dz
\end{equation}
where $z_{\text{SN}}$ and $\sigma_{\text{SN}}$ are the redshift and redshift uncertainty of the supernova (given by $\sigma_z$ above), $z_{\text{CL}}$ and $\sigma_\text{CL}$ are the redshift and redshift uncertainty of the cluster, and $z_d$ is three times the velocity dispersion of the cluster in redshift space. We adopt the cluster redshifts as tabulated in the MCXC catalog and set $\sigma_{\text{CL}} = 0$ as this uncertainty is negligible compared to $\sigma_{\text{SN}}$. 
We use the cluster scaling relation given by \cite{2011A&A...526A.105Z} to map the catalogued $r_{500}$ to a cluster velocity dispersion that is used to calculate $z_d$. We assume cluster membership for any supernova that yields $p > 0.5$.

In order to compare our cluster supernova samples to the field, we construct samples of SNe Ia in quiescent field galaxies and star-forming field galaxies. For our purposes, the ``field'' includes any $0.01 < z < 0.1$ galaxies outside of our rich x-ray clusters: we do not attempt to eliminate galaxies in groups, in poorer clusters (e.g., optically selected), or in otherwise overdense environments. To identify quiescent and star-forming field galaxies, we use the catalog from \cite{2015ApJS..219....8C}, which contains star formation rates (SFRs) and stellar masses for around 850,000 galaxies based on Sloan Digital Sky Survey (SDSS) + WISE photometry \citep{2000AJ....120.1579Y,2010AJ....140.1868W}. SN host galaxy associations for the field sample use the DLR method \citep{2006ApJ...648..868S, 2016AJ....152..154G}. We base our quiescent or star-forming classification on \cite{2015ApJS..219....8C}, except with slightly stricter criteria\footnote{For galaxies with $(r - z)_{\text{rest}} < 0.625$, we classify those with $(u - r)_{\text{rest}} \ge 2.1$ as quiescent, and those with $(u - r)_{\text{rest}} \le 1.9$ as star-forming. For galaxies with $(r - z)_{\text{rest}} \ge 0.625$, our quiescent galaxies have $(u - r)_{\text{rest}} \ge 1.6\,(r - z)_{\text{rest}} + 1.1$ and star-forming galaxies have $(u - r)_{\text{rest}} \le 1.6\,(r - z)_{\text{rest}} + 0.9$. See Figure 2 of \cite{2015ApJS..219....8C}.} to avoid the ambiguity of galaxies that lie in the ``green valley'' of star formation \citep{Salim:2014}.

While we can create large enough field supernova samples even restricting the sky area to the SDSS footprint, our cluster SNe~Ia cover the whole sky. For the cluster SN host galaxies with SDSS photometry \citep[just under half of the sample;][]{2015ApJS..219....8C}, we use the same quiescent/star-forming classification as the field galaxies above.\footnote{One supernova in our \tworadius sample, SN~2020jny, has a host that falls in the green valley between our SDSS color regions, so we do not include its host in either the cluster quiescent or cluster star-forming samples.} For the cluster host galaxies without SDSS photometry, we rely on WISE colors alone, classifying such galaxies as star-forming if $W2 - W3 = [4.6\mu] - [12 \mu] > 1.4$, and quiescent otherwise \citep[see Figure 12 of][]{2010AJ....140.1868W}. 

A few special cases were handled separately: SN~2019ulw (in our \tworadius sample) had an identifiable host, but lacked either WISE or SDSS photometry, so we exclude it from our cluster star-forming and quiescent samples. The hosts of SN~2007fr and SN~2019gwn (in our \radius sample) also lacked the requisite photometry, but we were able to acquire slit spectra centered on the host galaxy using the Robert Stobie Spectrograph \citep[RSS;][]{2006SPIE.6269E..2AS} on the Southern African Large Telescope. The spectral reductions were performed using a custom pipeline based on PySALT \citep{2010SPIE.7737E..25C}. The spectrum of SN~2007fr's host showed prominent narrow emission lines of H$\alpha$, H$\beta$, and [O III], so we categorize this galaxy as star-forming. The host galaxy of SN~2019gwn was faint and while its spectrum showed hints of an H$\alpha$ emission line, we could not confidently classify categorize it, and so we exclude SN~2019gwn from the star-forming or quiescent samples.
 
\subsection{Supernova photometry} \label{sec:photometric_data}

We use archival photometry for our cluster and field supernova samples. The bulk of these data is drawn from the Zwicky Transient Facility \citep[ZTF;][]{2019PASP..131a8002B} and the Asteroid Terrestrial-impact Last Alert System \citep[ATLAS;][]{2018PASP..130f4505T}. 
For the ZTF data we used the forced photometry service \citep{2019PASP..131a8003M} to obtain \emph{gri} magnitudes\footnote{ZTF \emph{i}-band data are initially proprietary, so we only used the \emph{i}-band photometry through mid-2021, publicly released in ZTF DR16.}. We also used the ATLAS forced photometry service \citep{2020PASP..132h5002S,2021TNSAN...7....1S} to gather \emph{oc} (the wide orange and cyan ATLAS passbands) supernova light curves. We also made extensive use of the ALeRCE broker \citep{2021AJ....161..242F} to examine light curves and compare photometric data. For supernovae with both ZTF and ATLAS photometry, we confirmed consistency in SN~Ia light curve fits (see \autoref{sec:params}) compared to ZTF data alone. Because we have a large field supernova sample, we restrict it to exclusively use photometry from ZTF, ATLAS, or both.

\begin{table*}
    \begin{center}
    \begin{tabular}{|c|c|c c c c|c c c c c|} 
    \hline
    \multicolumn{2}{|c|}{} & \multicolumn{4}{|c|}{unimodal \xone} & \multicolumn{5}{|c|}{bimodal \xone} \\
    \hline
     \hline
     Sample & $N_{\text{SN}}$ & Mean & Std Dev. & Median & MAD & $f_1$ & $\mu_1$ & $\sigma_1$ & $\mu_2$ & $\sigma_2$\\ [0.5ex]
     \hline\hline
     \radius & 54 & $-$1.49 & 1.14 & $-$1.82 & 0.53 & $0.76^{+0.06}_{-0.06}$ & $-2.05^{+0.09}_{-0.08}$ & $0.47^{+0.07}_{-0.06}$ & $+0.37^{+0.19}_{-0.19}$ & $0.58^{+0.22}_{-0.14}$ \\ 
     \hline
     \tworadius & 48 & $-$0.62 & 1.19 & $-$0.44 & 1.15 & $0.39^{+0.07}_{-0.07}$ & $-1.91^{+0.08}_{-0.08}$ & $0.32^{+0.08}_{-0.06}$ & $+0.26^{+0.12}_{-0.12}$ & $0.61^{+0.12}_{-0.09}$ \\
     \hline
     full cluster & 102 & $-$1.08 & 1.24 & $-$1.53 & 0.91 & $0.59^{+0.05}_{-0.05}$ & $-2.01^{+0.06}_{-0.06}$ & $0.43^{+0.05}_{-0.05}$ & $+0.28^{+0.10}_{-0.11}$ & $0.60^{+0.10}_{-0.08}$ \\
     \hline
     field quiescent & 346 & $-0$.76 & 1.13 & $-$0.85 & 0.94 & $0.32^{+0.12}_{-0.09}$ & $-1.93^{+0.19}_{-0.15}$ & $0.47^{+0.11}_{-0.10}$ & $-0.22^{+0.23}_{-0.17}$ & $0.87^{+0.09}_{-0.12}$ \\ 
     \hline
     field star-forming & 395 & $+$0.20 & 0.97 & $+$0.27 & 0.56 & $0.07^{+0.03}_{-0.02}$ & $-1.91^{+0.33}_{-0.19}$ & $0.46^{+0.22}_{-0.13}$ & $+0.35^{+0.06}_{-0.05}$ & $0.77^{+0.05}_{-0.04}$\\
     \hline
     \hline
     \radius\ quiescent & 43 & $-$1.77 & 0.92 & $-$2.04 & 0.43 & $0.86^{+0.05}_{-0.06}$ & $-2.07^{+0.09}_{-0.09}$ & $0.49^{+0.08}_{-0.06}$ & $+0.33^{+0.23}_{-0.20}$ & $0.49^{+0.36}_{-0.19}$ \\
     \hline
     \tworadius\ quiescent & 28 & $-$0.80 & 1.06 & $-$0.54 & 1.08 & $0.43^{+0.10}_{-0.10}$ & $-1.89^{+0.10}_{-0.09}$ & $0.27^{+0.10}_{-0.06}$ & $+0.12^{+0.12}_{-0.08}$ & $0.54^{+0.16}_{-0.10}$ \\
     \hline
     full cluster\ quiescent & 71 & $-$1.39 & 1.09 & $-$1.76 & 0.60 & $0.69^{+0.06}_{-0.06}$ & $-2.03^{+0.07}_{-0.07}$ & $0.44^{+0.06}_{-0.05}$ & $+0.12^{+0.11}_{-0.08}$ & $0.54^{+0.13}_{-0.09}$ \\
     \hline
    \end{tabular}
    \end{center}
\caption{Number of supernovae for each sample, as well as the mean, standard deviation, median, median absolute deviation for the SALT3 \xone parameter distributions for each sample (unimodal) and double Gaussian fits (bimodal) to the \xone distribution, with $f_1$ indicating the fraction in the fast-declining population.}
\label{table:distribution_values}
\end{table*}

Our cluster samples included 10 supernovae with Pan-STARRS1 \citep{2012ApJ...750...99T} photometric data from the Young Supernova Experiment \citep[YSE;][]{Jones:2021} first light-curve data release \citep{2023ApJS..266....9A}. We included these \emph{gri} photometry in our analysis (but we did not include the \emph{z}-band). For cluster SNe~Ia that predated these surveys (before 2016), we retrieved available Johnson-Cousins \emph{BVRI} \citep{1990PASP..102.1181B} and SDSS \emph{gri} \citep{2000AJ....120.1579Y} photometry from varied sources via the Open Supernova Catalog \citep{2017ApJ...835...64G}. 

\begin{figure}
    \centering
    \includegraphics[scale=0.57]{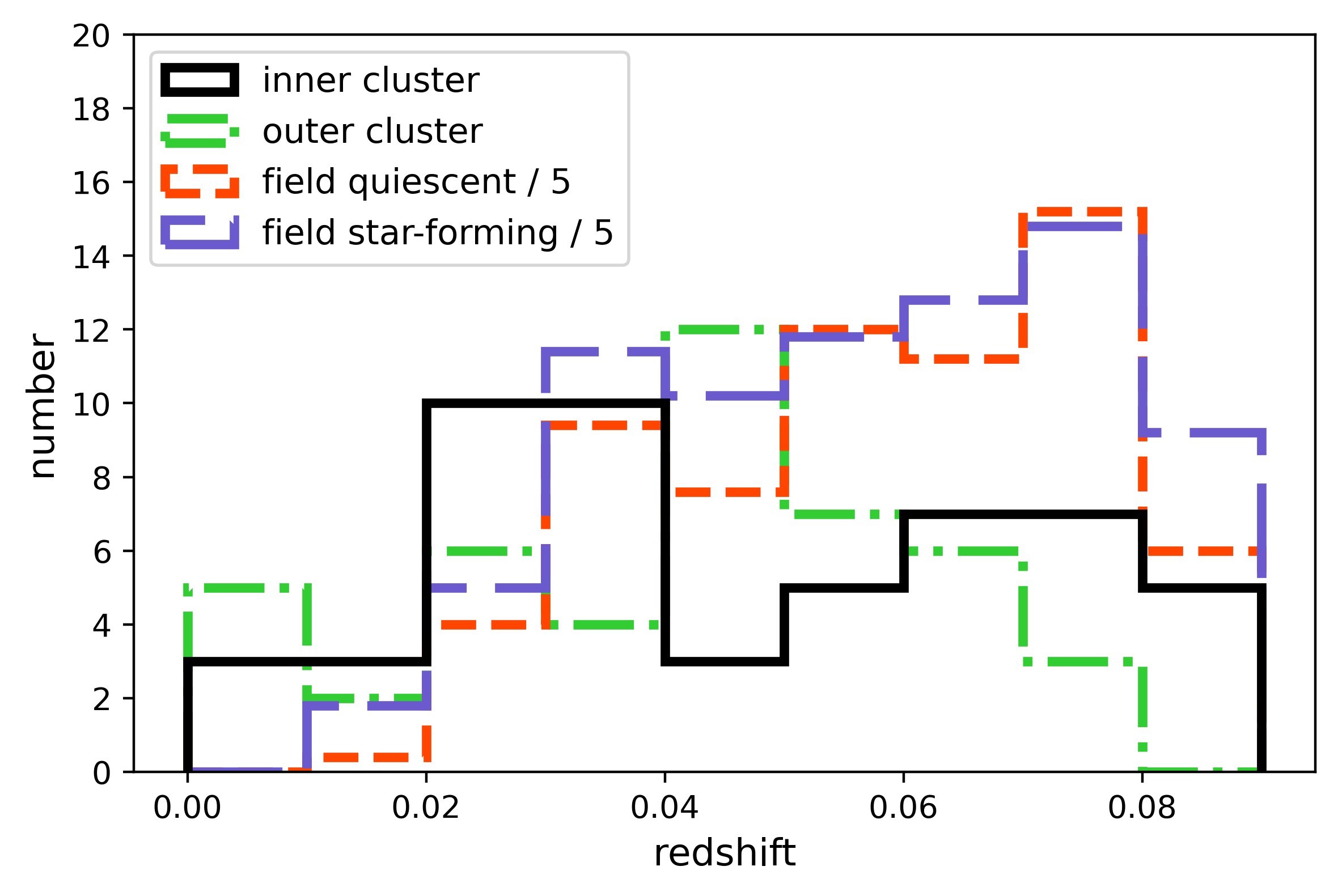}
    \caption{Redshift histogram for SNe in our cluster and field samples. The cluster sample redshift values are the host cluster redshifts, while the field sample redshifts are from the host galaxy or the supernova. The counts for the field samples have been divided by 5 to bring them on the same scale as the smaller cluster samples.}
    \label{fig:redshifts}
\end{figure}

\subsection{Supernova light curve fitting} \label{sec:params}

We employ the SALT3 model to fit our SN~Ia light curves \citep{2007A&A...466...11G,2021ApJ...923..265K}, combined with \cite{1998A&A...331..815T} standardization, using the SNCosmo package \citep{2016ascl.soft11017B}. Recent work has shown that the switch to SALT3 over SALT2 causes negligible difference in cosmological parameter estimation but reduces calibration errors \citep{2023MNRAS.tmp..345T}. SALT3 fits a multicolor SN~Ia light curve with three parameters: $x_0$, which captures the peak flux in the \emph{B} band; $x_1$, which parameterizes the light-curve decline (and rise) rate; and $c$, which measures the supernova color (corresponding approximately to \emph{B}$-$\emph{V}). A smaller $x_1$ indicates a faster-evolving light curve and a larger \textit{c} denotes a redder color.

From the SALT3 fits we can define a peak \emph{B} magnitude 
\begin{equation}
m_{B} = -2.5\,\log \left( x_0 \right) + 10.5
\end{equation}
where by convention $m_B = 10.5$ corresponds to $x_0 = 1$ \citep{2021ApJ...923..265K}. We can then derive a standardized magnitude and distance modulus with a light-curve width and color correction: 
\begin{equation}
\mu_\text{obs} = m_{B} + \alpha\,x_1 - \beta\,c - M_{B}
\end{equation}
where $\mu_\text{obs}$ represents the inferred distance modulus, and 
$\alpha$, $\beta$, and $M_{B}$ are fit parameters that we describe in \autoref{sec:cosmology}. 
        
We exclude any SNe that have less than five photometric measurements in total. We correct for effects of Milky Way dust extinction in our SALT3 model fits, with an assumed Milky Way $R_V = 3.1$ and $E(B - V)$ values along the line of sight to our SNe from the dust maps of \cite{1998ApJ...500..525S}, recalibrated in \cite{2011ApJ...737..103S}. We make use of the NED extinction calculator tool through an existing Python script.\footnote{\url{https://github.com/mmechtley/ned_extinction_calc}}

To create our final supernova samples, we apply light curve quality and fit parameter cuts. As is typical in cosmological analyses, we require SALT3 fits with $|x_1| < 3.0$ and $|c| < 0.3$, and uncertainties $\sigma(x_1) < 1.0$ and $\sigma(c) < 0.2$. We also require a fit uncertainty on the time of maximum light $\sigma(t_0) < 0.5$ days. Similar cuts were introduced by \citet{2014A&A...568A..22B} and have been used in other SN~Ia cosmological studies \citep{2018ApJ...857...51J,2019ApJ...881...19J,2019ApJ...874..150B,2021ApJ...909...26B,2022ApJ...938..112P}. For our cluster samples, we further manually inspect the light curve fits and demand that the light curves have both pre-maximum and post-maximum data. 

\section{Results} \label{sec:results}

\subsection{Light-Curve Properties} \label{sec:sample_comparison}

\begin{figure*}
    \centering
    \includegraphics[scale=0.9]{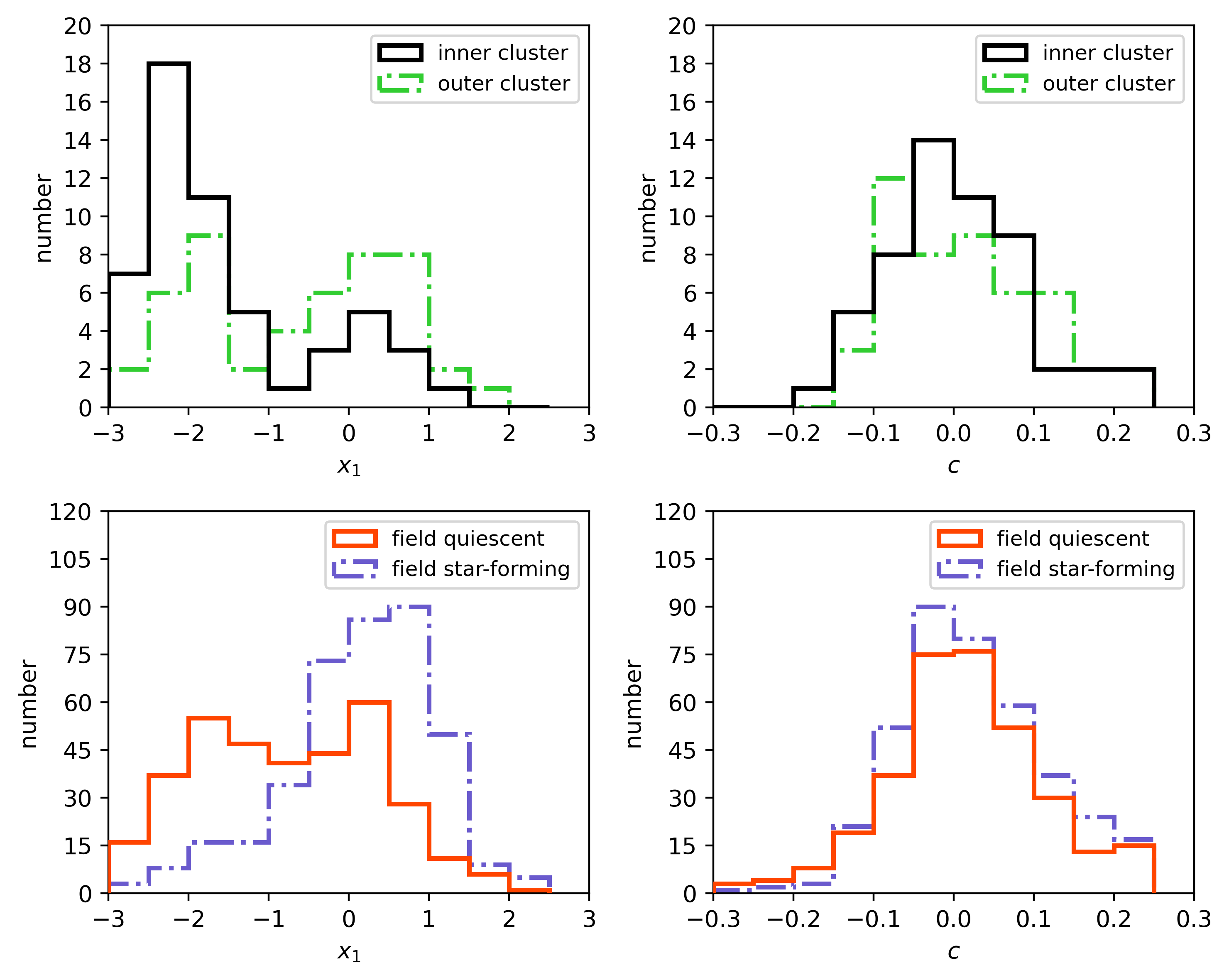}
    \caption{\textit{Top:} Histograms of the full SALT3 $x_1$ and $c$ parameter distributions for our cluster SN~Ia samples. \textit{Bottom:} Histograms showing the same distributions as above, but for our field quiescent and field star-forming samples.}
    \label{fig:param_comparison}
\end{figure*}

In our \radius sample, we have 54 SNe Ia with adequate light curves and SALT3 parameter values that fall within our cutoff ranges and our \tworadius sample contains 48 SNe that pass these cuts. Our field quiescent and field star-forming samples have 346 and 395 SNe, respectively, that pass the cuts. \autoref{fig:redshifts} shows a histogram of the redshifts for these SN~Ia samples. The median redshifts for the field quiescent, field star-forming, \radius, and \tworadius samples are: 0.062, 0.060, 0.044, and 0.045, respectively. The lower median redshifts of the cluster samples is likely a result of the cluster redshift distribution in the flux-limited x-ray selection for the MCXC catalog \citep{2011A&A...534A.109P}. We explore potential effects of the slightly different redshift distributions below. The cosmic age difference between the field quiescent sample median redshift ($z = 0.062$) and the \radius sample median ($ z = 0.044$) is about 230 Myr for our adopted cosmology.

\autoref{fig:param_comparison} shows the distributions of the SALT3 \xone and $c$ parameters for our \radius and \tworadius samples, compared with the field quiescent and field star-forming samples. The most striking differences are seen in \xone. In SALT2 (and SALT3) model training this light-curve shape parameter is created to have zero mean and unit standard deviation \citep{2007A&A...466...11G,2021ApJ...923..265K} across the training set. However, here we see a strong environmental dependence in the \xone distribution. A statistical summary of these distributions is given in the ``unimodal'' columns of \autoref{table:distribution_values}. While the field star-forming sample is not far from a mean of zero and standard deviation of one, the other samples are markedly different. Fast-declining (lower \xone) SNe~Ia have long been known to preferentially occur in quiescent galaxies \citep{1996AJ....112.2391H,2000AJ....120.1479H,1996ApJ...465...73B}, and this is borne out comparing our field quiescent and field star-forming samples.
Moreover, fast-declining SNe~Ia \emph{dominate} the \radius sample, where the \xone distribution is strongly peaked approximately two standard deviations lower than the mean of the training data. To be clear, when we refer to ``fast-declining" SNe~Ia, we mean any SNe~Ia with low \xone values, not just traditionally fast-evolving SN~Ia sub-types.

\begin{figure}
    \centering
    \includegraphics[scale=0.58]{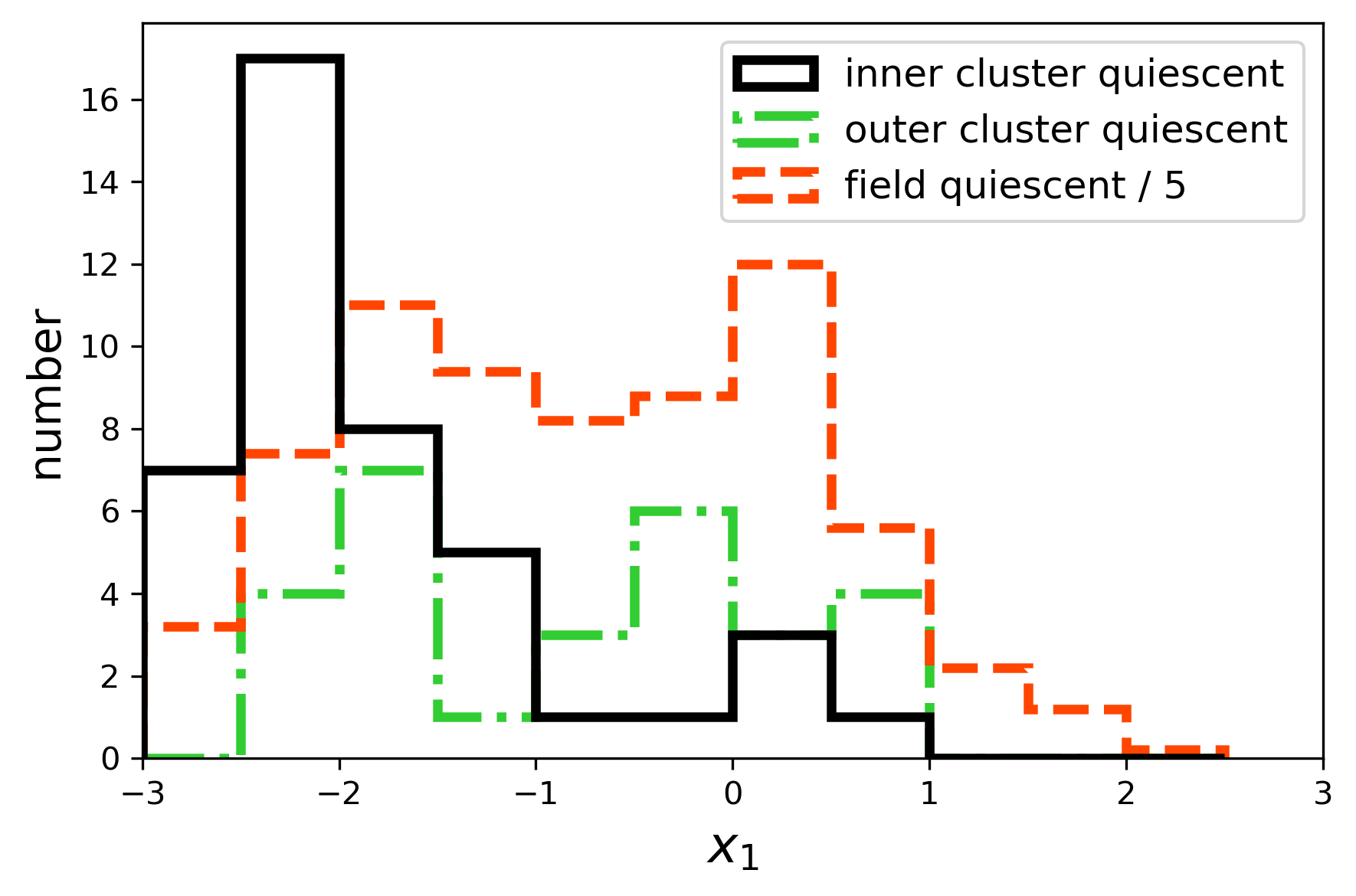}
    \caption{Histogram of \xone parameter values for SNe in our \radius and \tworadius samples, restricted to quiescent host galaxies. These are compared to the field quiescent host galaxy sample (whose counts are scaled down by 5 to ease comparison).}
    \label{fig:quiescent_comparison}
\end{figure}

To better understand the environmental dependence in supernova properties, it is useful to control for the host galaxy type. In \autoref{fig:quiescent_comparison} we limit the cluster samples to comprise quiescent host galaxies only and compare these with the field quiescent sample. Our \radius and \tworadius quiescent samples consist of 43 SNe Ia and 28 SNe Ia, respectively. The \tworadius quiescent \xone distribution is similar to the field quiescent one, whereas the \radius quiescent distribution is even more strongly peaked with fast-declining SNe.

To quantitatively compare these distributions, we employ a two-sample Anderson-Darling (A-D) test, which tests the null hypothesis whether two empirical samples are drawn from the same distribution \citep{Pettitt1976ATA}. Calculating the test statistic between our \radius and \tworadius quiescent \xone samples, we find $p < 0.001$, indicating a clear difference in these populations. We similarly find $p < 0.001$ for the \radius quiescent and field quiescent samples, but for the \tworadius quiescent and field quiescent samples we do not find evidence for different \xone distributions, $p > 0.25$. This suggests the \radius sample is the standout among quiescent host galaxies.

Performing a similar analysis for star-forming host galaxies is hampered by small number statistics. Our radius and \tworadius star-forming samples consist of only 8 and 15 SNe Ia, respectively. If we combined these to form a cluster star-forming host sample, we find $p = 0.031$ for the A-D test between the cluster star-forming and field star-forming \xone distributions. There is thus only marginal evidence for a population difference between cluster and field SNe~Ia in star-forming hosts.

In contrast to the \xone distributions, the right panels of \autoref{fig:param_comparison} show relatively similar SALT3 $c$ across all of our cluster and field samples. A-D tests confirm this impression; we find no evidence for significant population differences in the color distributions of these SNe.

Distinct from the field star-forming sample, the \xone distributions for the quiescent hosts (field or cluster) are bimodal, supported by sample statistics. Two-population models for SNe~Ia, driven by their \xone distributions, have been explored before \cite[e.g., recently by][who fit such a model to a full sample of SNe~Ia; see \autoref{sec:discussion}]{2023MNRAS.525.5187W}, but isolating objects in quiescent hosts (and especially our \radius sample), brings out the bimodality clearly. We investigate a two-population \xone distribution by running a Markov-Chain Monte Carlo (MCMC) fit to a double Gaussian model. The fit parameters are $\mu_1$ and $\mu_2$, the \xone means of faster and slower declining populations, respectively; $\sigma_1$ and $\sigma_2$, the widths of the two populations; and $f_1$, the fraction of the sample in the faster-declining population (so that the fraction of the slower-declining population is $1 - f_1$).  

The results of these fits are summarized in \autoref{table:distribution_values}. In all samples we find a fast-declining population centered at $x_1 \simeq -2$ that is narrower in width\footnote{We note that part of this narrower width may be ascribed to the truncation at $x_1 > -3$.} than a broader, slower-declining population centered at $x_1 \simeq +0.3$, with slight variations between samples. There is a strong environmental variation in the fraction of objects in the fast-declining population, from approximately 76\% in the inner cluster sample (and 86\% if we restrict to inner cluster quiescent hosts) all the way down to just 7\% in the field star-forming sample.

\begin{figure}
    \centering
    \includegraphics[scale=0.46]{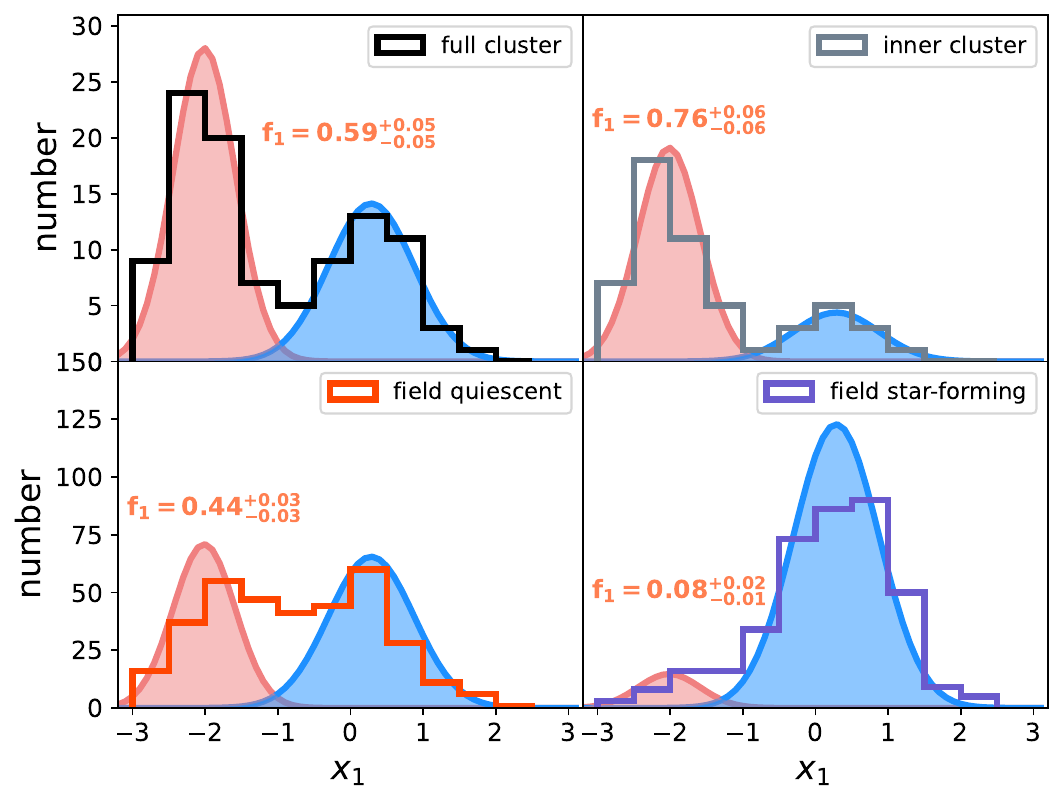}
    \caption{Bimodal fits to the \xone distributions in different samples. Here we fix the double Gaussian parameters ($\mu_1$, $\sigma_1$, $\mu_2$, $\sigma_2$) to fit the full cluster sample (upper left). The fast-declining population is shown in coral and the slower-declining population is shown in blue. Using this fixed model, in each of the other panels we fit only $f_1$, the fraction of objects in the fast-declining population, for the \radius (upper right), field quiescent (lower left), and field star-forming (lower right) samples.}
    \label{fig:bimodal_fits}
\end{figure}

We illustrate these results visually in \autoref{fig:bimodal_fits}. Different than in \autoref{table:distribution_values}, in \autoref{fig:bimodal_fits}, we fix the two population Gaussians ($\mu_1$, $\sigma_1$, $\mu_2$, $\sigma_2$) as fit to the \emph{full cluster} sample (upper left panel). Then for the other three samples displayed (inner cluster, field quiescent, and field star-forming), we only re-fit for $f_1$, to better isolate the changing fraction of fast-declining supernovae.  We obtain $f_1$ values of $76 \pm 6\%$, $44\% \pm 3\%$, and $8 \pm 2\%$ for these three samples, respectively. Not only is there a vast difference compared to the field star-forming sample, there is even a nearly 5$\sigma$ difference in $f_1$ between the \radius and field quiescent samples. Clearly, the \radius environment produces a different population of SNe~Ia than would be predicted for similar host galaxies in the field.

Some of the differences between the full double Gaussian fits in \autoref{table:distribution_values} and the model fixed to the full cluster sample can also be seen in \autoref{fig:bimodal_fits}. The \radius sample fast-declining population is even slightly faster than the full cluster sample. The peaks in the field quiescent data are broader and not as well separated as in the full cluster sample, and the slower-declining peak in the field star-forming sample is also somewhat broader than the corresponding population in the cluster samples. 

\begin{figure}
    \centering
    \includegraphics[scale=0.6]{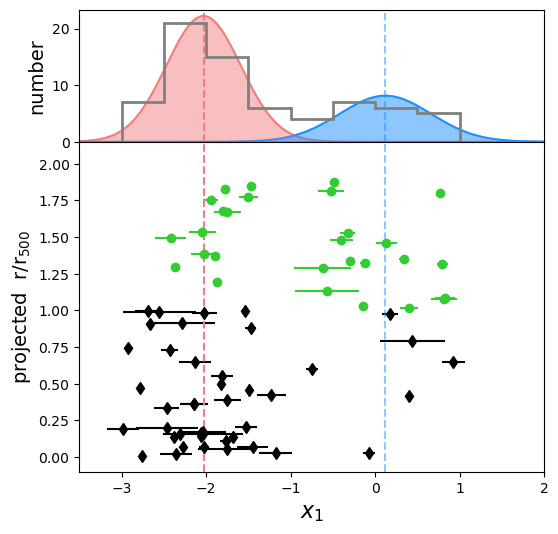}
    \caption{\textit{Top:} histogram of our full quiescent cluster sample in gray, with the fit bimodal populations overlayed in the coral and blue colors respectively. The vertical dashed lines represent the fit means for each population distribution for this sample. \textit{Bottom:} the \xone parameter values of our cluster SNe Ia in quiescent hosts as a function of their projected distance from their cluster center. The \radius portion of the sample is represented by black diamonds. The \tworadius portion of the sample is represented by green circles. We can see that SNe closer to the center of the cluster tend to have much faster-evolving light curves.}
    \label{fig:cluster_sep}
\end{figure}

The difference in the \xone distribution between our \radius SNe~Ia sample and the \tworadius leads us to examine how this varies as a function of the projected distance from the center of the cluster. In \autoref{fig:cluster_sep}, we see a clear paucity of slowly-declining SNe~Ia near the cluster centers. There is also a hint that the fast-declining population may become slightly slower-declining in the outskirts of the clusters. 

\begin{table*}
    \begin{center}
    \begin{tabular}{c|c c c c c} 
     \hline
     Sample & $\alpha$ & $\beta$ & $M_B$ (mag) & $\sigma_{\text{int}}$ (mag) & RMS (mag) \\ [0.5ex] 
     \hline\hline
     \radius & $0.133^{+0.021}_{-0.021}$ & $2.381^{+0.299}_{-0.304}$ & $-19.247^{+0.039}_{-0.039}$ & $0.154^{+0.022}_{-0.018}$ & 0.163\\ 
     \hline
     \tworadius & $0.128^{+0.023}_{-0.024}$ & $2.524^{+0.308}_{-0.302}$ & $-19.343^{+0.032}_{-0.032}$ & $0.164^{+0.023}_{-0.019}$ & 0.163\\ 
     \hline
     field quiescent & $0.159^{+0.008}_{-0.008}$ & $2.135^{+0.083}_{-0.082}$ & $-19.311^{+0.011}_{-0.010}$ & $0.141^{+0.007}_{-0.006}$ & 0.150\\ 
     \hline
     field quiescent ($z < 0.06$) & $0.157^{+0.014}_{-0.014}$ & $2.447^{+0.128}_{-0.130}$ & $-19.284^{+0.022}_{-0.022}$ & $0.164^{+0.011}_{-0.010}$ & 0.172\\ 
     \hline
     field star-forming & $0.113^{+0.008}_{-0.008}$ & $2.608^{+0.079}_{-0.079}$ & $-19.254^{+0.007}_{-0.007}$ & $0.127^{+0.006}_{-0.005}$ & 0.134\\ 
     \hline
     full field & $0.127^{+0.005}_{-0.005}$ & $2.395^{+0.059}_{-0.058}$ & $-19.266^{+0.006}_{-0.006}$ & $0.137^{+0.004}_{-0.004}$ & 0.144\\ 
     \hline
    \end{tabular}
    \end{center}
\caption{Fit parameters obtained through our cosmological MCMC procedure for each sample.}
\label{table:mcmc_values}
\end{table*}

\begin{figure}
    \centering
    \includegraphics[scale=0.58]{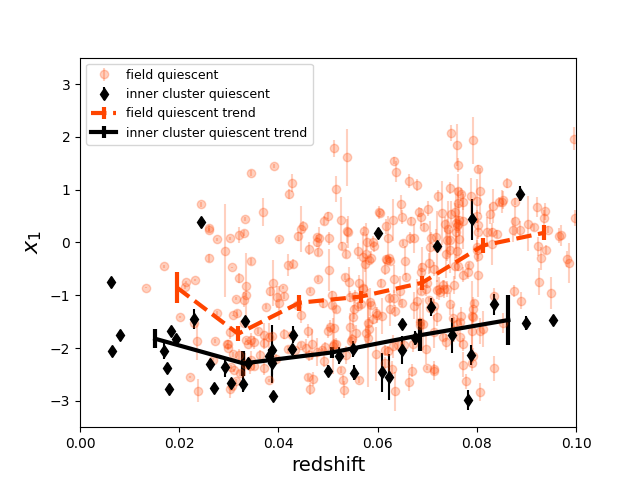}
    \caption{Trends in the \xone parameter values of our field quiescent and \radius quiescent samples as a function of redshift. The black and orange points are the binned median values for our \radius quiescent and field quiescent samples respectively. The errors on each point are the standard error for the bin. The points are positioned at the centers of each bin.}
    \label{fig:redshift_trend}
\end{figure}

We note that we only observe the projected separation of the supernova and its host galaxy within the cluster, so some of our \radius\ sample objects may physically be part of our \tworadius sample (or even further out). We use a cluster galaxy number density model \citep{1997ApJ...485L..13C} to estimate that up to $\sim$28\% of our \radius sample could be contaminants. For 54 total \radius SNe~Ia, this means up to $\sim$15 could be projected from further out. If those objects follow the \tworadius \xone distribution (\autoref{table:distribution_values}), approximately 61\% (9) of those should be from the slowly-declining population, with about 6 in the fast-declining population. Subtracting these out of our \radius sample would leave 41 $-$ 6 = 35 fast-decliners ($\xone < -1$) and just 13 $-$ 9 = 4 slower-decliners ($\xone > -1$), corresponding to a projection-corrected \radius $f_1$ approaching 90\% (35 out of 39)! If we apply a similar projection correction for just the \radius quiescent galaxies, we would find that it could account for all of the slow-decliners in the sample. While these would imply an extreme population skew for \radius SNe~Ia, we can rule out the possibility that \emph{all} \radius objects are fast-decliners: SN~2018bgs is in the brightest cluster galaxy (BCG) and is in the slower-declining population. We further caution that our \radius and \tworadius separation is a simplification based on assuming a spherical geometry can adequately describe the clusters.

Though we have constructed a nearby, $z < 0.1$, sample, we can still investigate trends with redshift. In \autoref{fig:redshift_trend} we show the \radius quiescent sample \xone distribution as a function of redshift, comparing it to the field quiescent sample. Note that at redshifts $z < 0.06$, there is only one SN with \xone $>$ 0 in the \radius quiescent sample: SN~2008bf. 
All of the other slower-declining SNe in this sample are at higher redshifts. 

Both the \radius quiescent sample and the field quiescent sample in \autoref{fig:redshift_trend} show a trend towards larger \xone as redshift increases, even in the faster-declining population. This could be a result of Malmquist bias, as slower-evolving SNe Ia tend to be more luminous before standardization. Such a luminosity bias could not be used to explain the lack of slower-declining SNe~Ia at low redshift in the \radius quiescent sample, however, as these brighter SNe should be most easily detected, and they are clearly present in the field quiescent sample.

\subsection{Standardization and Cosmological Distances} \label{sec:cosmology}

We now turn our attention to examining whether these environmental differences among the samples persist through SN~Ia standardization and inferred distances for cosmology. As mentioned in \autoref{sec:params}, in order to do a cosmological analysis with the \citet{1998A&A...331..815T} standardization, we need to fit for the fit parameters $\alpha$, $\beta$, and $M_{B}$. We also fit for $\sigma_{\text{int}}$, a measure of the intrinsic scatter that exists within our SN samples. For our fits, we use an MCMC implemented through the \texttt{emcee} package \citep{2013PASP..125..306F}. Our log-likelihood function, $\ln \mathcal{L}$, is defined via the relation
\begin{equation}
    -2 \ln \mathcal{L} = \sum\limits_i \ln \left(2\pi[\sigma_{\text{obs},i}^2+\sigma_{\text{int}}^2]\right) + \cfrac{(\mu_{\text{obs},i}-\mu_{\text{cosmo},i})^2}{\sigma_{\text{obs},i}^2 + \sigma_{\text{int}}^2},
\end{equation}
where $\mu_{\text{cosmo}}$ is the distance modulus derived from our assumed cosmology,
\begin{gather}
    \mu_{\text{cosmo}} = 5\log \left(d_L/\text{Mpc}\right) + 25\\[10pt]
    d_L = \frac{c (1+z)}{H_0}\int_{0}^{z}\frac{dz'}{\sqrt{ \Omega_{M}(1+z')^3+\Omega_{\Lambda}}},
\end{gather}
and $\sigma_{\text{obs}}$ is the distance modulus uncertainty for each SN. This uncertainty comprises the variances and covariances of the SALT3 fit parameters, redshift uncertainty, and a negligible contribution from lensing effects \citep[given our low-redshift sample;][]{2010MNRAS.405..535J}.

\begin{figure}
    \centering
    \includegraphics[scale=0.35]{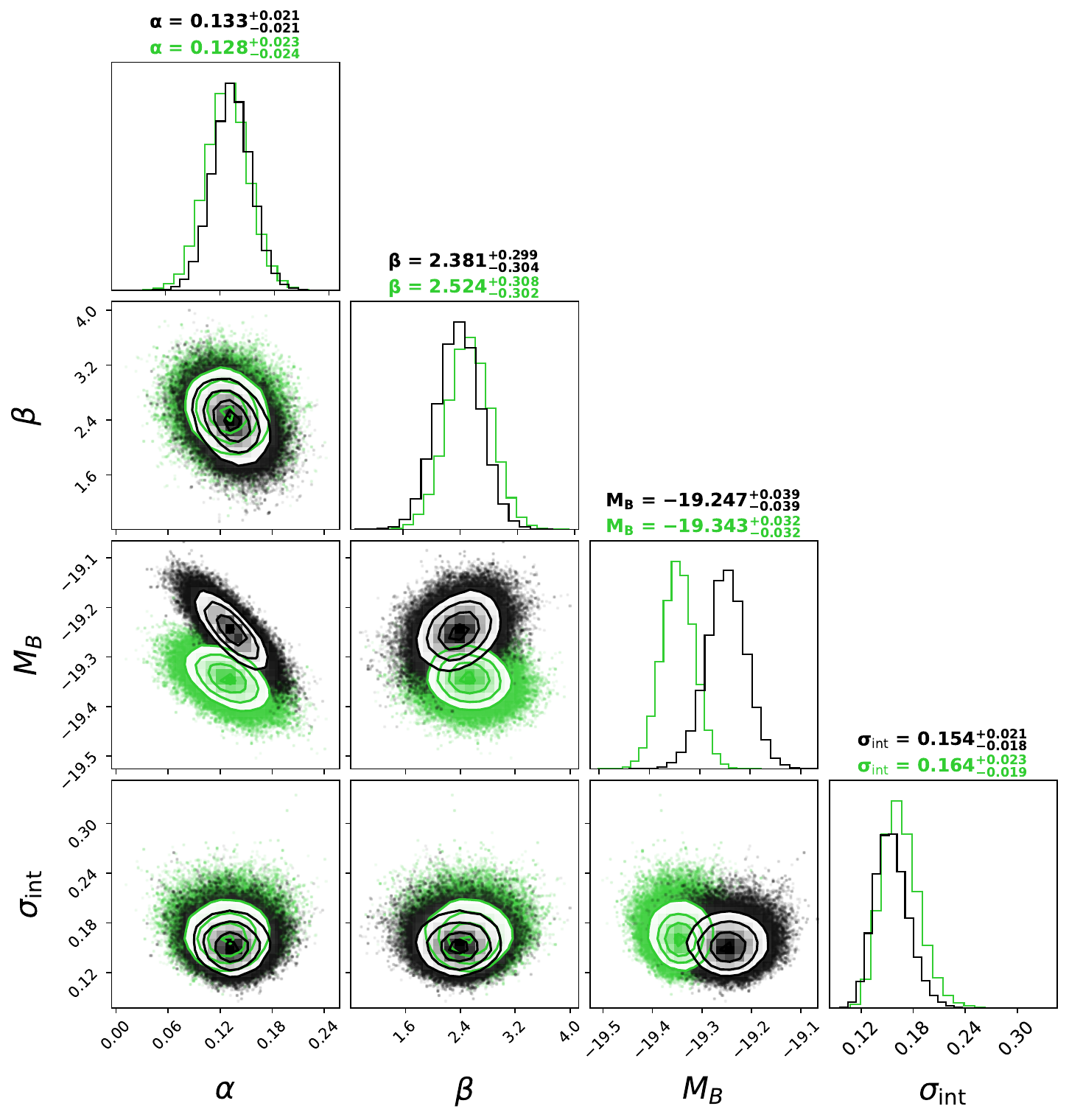}
    \caption{Corner plot for our \radius (black) and \tworadius (green) sample fit parameters. The best fit values and errors are on top of each corresponding column, with the \radius sample values on top and the \tworadius values below them. $M_B$ and $\sigma_{\mathrm{int}}$ are in units of magnitude.}
    \label{fig:corner_plot}
\end{figure}

For our cluster samples, we use the cluster redshifts as the cosmological redshifts, converted to the cosmic microwave background (CMB) frame. We make no correction for cluster peculiar velocities and include a 300~km~s$^{-1}$ peculiar velocity contribution to the redshift uncertainty \citep{2018A&A...615A.162L}, and further restrict the sample to $z > 0.01$. For our field samples, we convert host redshifts to the CMB frame and also correct for peculiar velocities, following \cite{2022ApJ...938..112P} and \cite{2022PASA...39...46C}, and using the velocity fields of \cite{2015MNRAS.450..317C} and \cite{2020MNRAS.497.1275S}. We assume a peculiar velocity uncertainty of 150~km~s$^{-1}$ for our field objects.

For the fit parameters, we adopt uniform priors on $\alpha$, $\beta$, and $M_B$, and a logarithmic prior on $\sigma_{\text{int}}$ with $\hat{p} \propto 1/\sigma_{\text{int}}$. We iterate our fit twice, removing 2$\sigma$ outliers in Hubble residual ($\mu_{\text{obs}} - \mu_{\text{cosmo}}$) after the first pass and rerunning to obtain our final values. The results of this analysis for all of our samples are summarized in \autoref{table:mcmc_values}. 

In \autoref{fig:corner_plot}, we show the corner plot for the \radius and \tworadius distributions from our MCMC analysis. The results between the two samples are largely consistent, though there is a hint of a 1.5$\sigma$ offset in $M_B$: $0.083 \pm 0.053$ mag. The \radius sample and, to a lesser extent, the \tworadius sample also show covariance between $M_B$ and $\alpha$ that can largely be ascribed to the \xone distributions in these samples. For the \radius sample especially, the average \xone is far from \xone $= 0$ that defines $M_B$, inducing a correlation with the slope $\alpha$.

\begin{figure}
    \centering
    \includegraphics[scale=0.48]{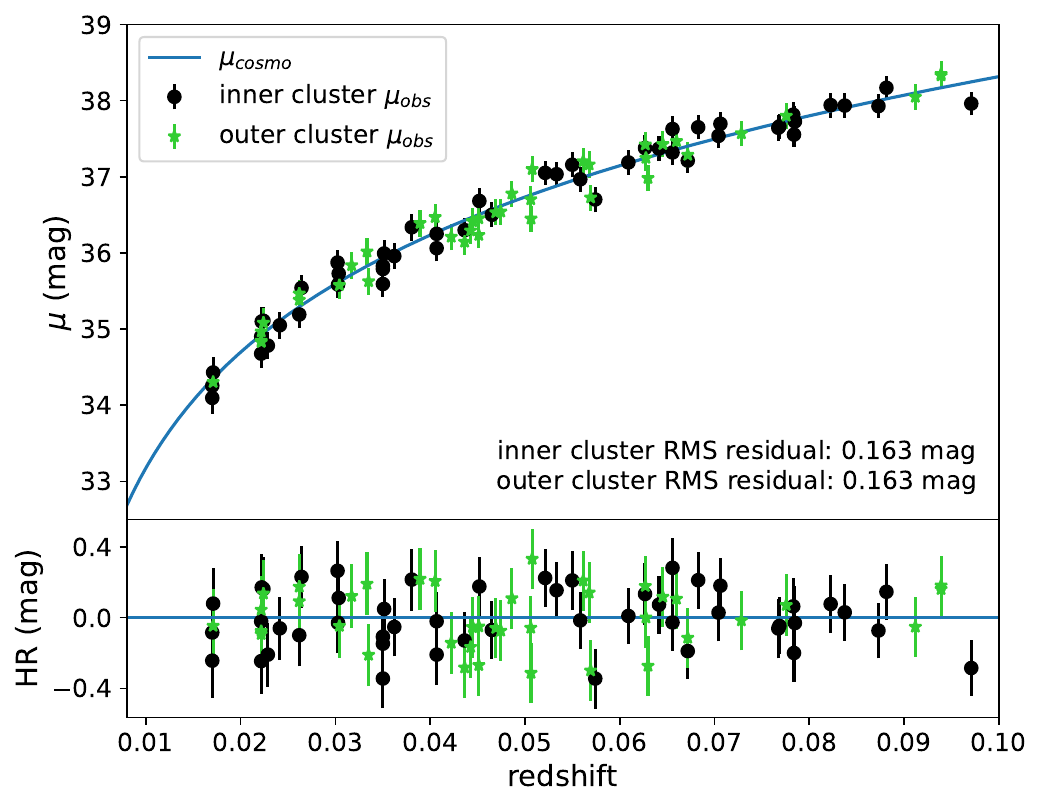}
    \caption{Inferred distances for the \radius and \tworadius samples, adopting best-fit values for the parameters $\alpha$, $\beta$, $M_B$, and $\sigma_\text{int}$ for each sample separately. The blue curve shows the predicted distance moduli from our assumed cosmology. The bottom panel shows the Hubble residuals, $\mu_\text{obs} - \mu_\text{cosmo}$. The points represent the samples after the 2$\sigma$ outlier removal.}
    \label{fig:moduli}
\end{figure}

In \autoref{fig:moduli}, we show the inferred distances to our cluster SNe~Ia compared to the assumed cosmological model. The SN distance moduli and their uncertainties depend upon the fit parameters ($\alpha$, $\beta$, $M_{B}$, and $\sigma_{int}$) and the individual SN light-curve parameters. The redshifts are taken to be the CMB-frame cluster redshifts. Though the \radius and \tworadius samples have different light curve properties, there are not major differences in the inferred distances. Both cluster samples give residual RMS of approximately $0.17$ mag, matching, for example, the RMS seen in a ZTF sample \citep{2022MNRAS.510.2228D}. 

Comparing our cosmological fits from the cluster samples to the field samples, we see in \autoref{table:mcmc_values} the field sample RMS is slightly lower than the cluster SNe~Ia, with RMS of 0.144 mag for the field quiescent sample and 0.134 mag for the field star-forming samples. Based on 
\autoref{fig:redshift_trend}, we noted the possibility of Malmquist bias affecting the field quiescent sample at $z > 0.06$. If we restrict the field quiescent sample to $z < 0.06$, \autoref{table:mcmc_values} shows a higher RMS residual of 0.172 mag, comparable to the cluster samples.

\begin{figure}
    \centering
    \includegraphics[scale=0.59]{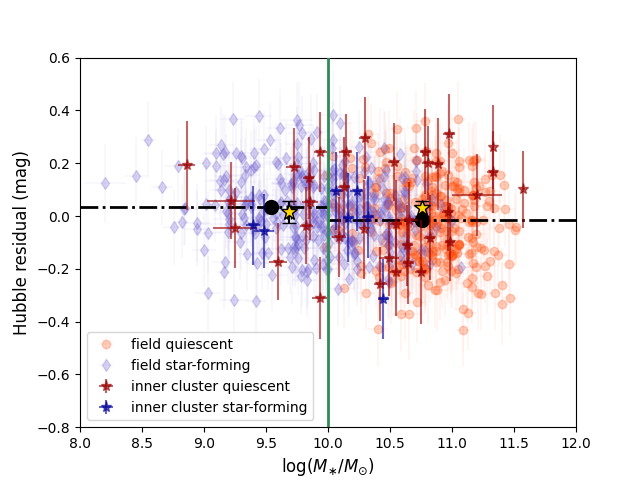}
    \caption{Hubble residuals for our field quiescent and star-forming host samples versus host-galaxy stellar mass. The green line represents our adopted dividing line of 10$^{10} \; M_\odot$ for the low-mass and high-mass samples. The black points are the weighted average of the mass bins, while the gold stars with the black outlines represent the weighted average of our full \radius sample in the lower and upper mass bins}. Outliers beyond 2$\sigma$ have been removed from the sample.
    \label{fig:mass_step}
\end{figure}

\begin{figure*}
    \centering
    \includegraphics[scale=0.6]{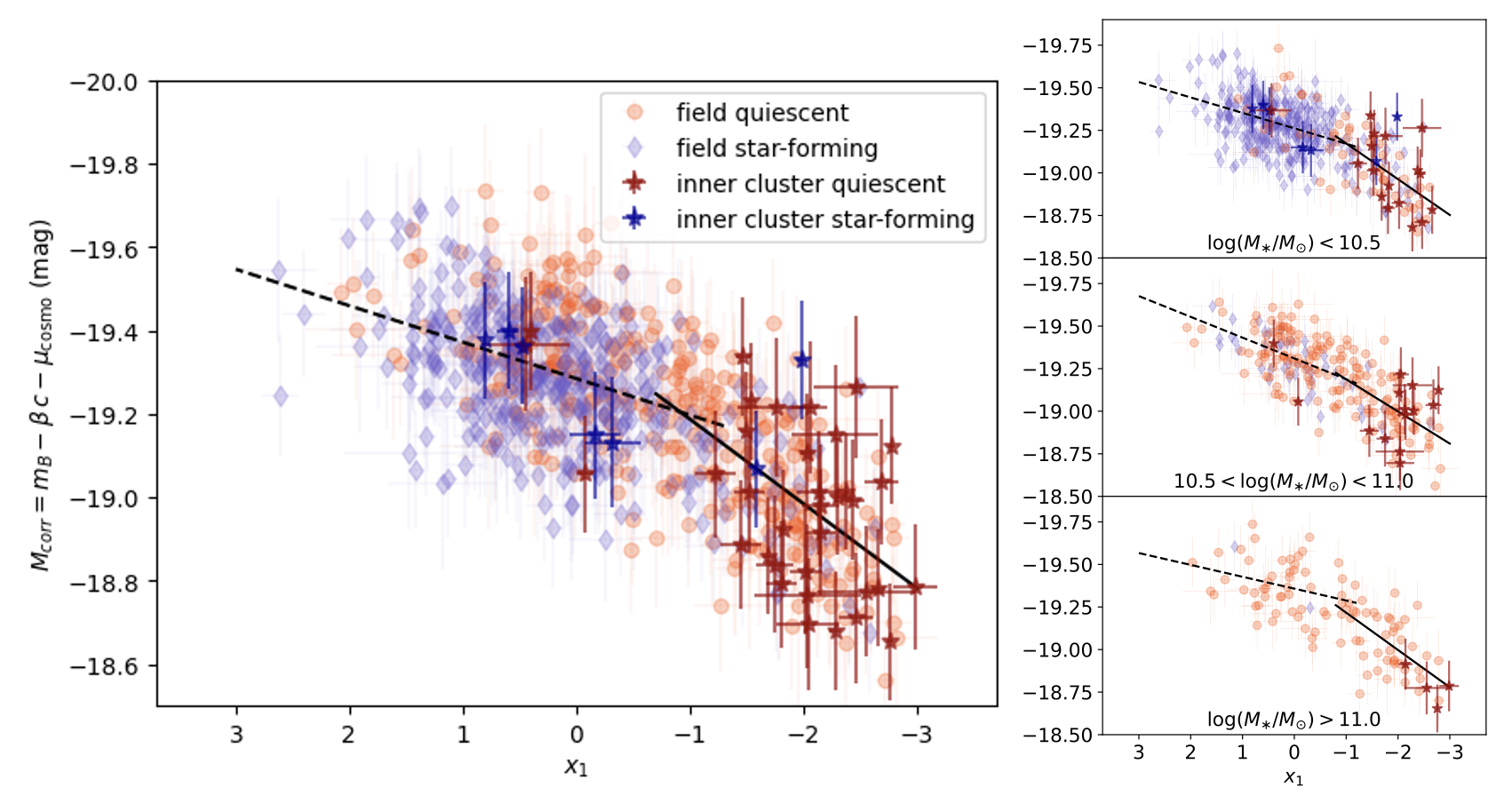}
    \caption{Color-corrected absolute magnitudes of field quiescent, field star-forming, inner cluster quiescent, and inner cluster star-forming host SNe Ia versus \xone. The black lines show weighted linear regression fits to the slower-declining (\xone > $-$1) and faster-declining (\xone < $-$1) SNe Ia from the four samples above. Confirming the result of \citet{2023ApJ...953...35G}, there is a clear steepening in the slope, representing the parameter $\alpha$, towards faster-declining SNe Ia. On the right, we show the same plot but for three increasing host stellar mass bins. We find this steepening in $\alpha$ to be consistent across all three mass bins.}
    \label{fig:alpha_fits}
\end{figure*}

The small sizes of our cluster samples mean that the best-fit parameters $\alpha$, $\beta$, and $M_B$ are uncertain enough to be consistent with both the field quiescent and field star-forming samples. However, the best-fit values diverge between the field quiescent and star-forming samples themselves, with $\sim$3$\sigma$ differences for each parameter. The field host star-formation rate is highly correlated with host stellar mass. In \autoref{fig:mass_step} we show the Hubble residual for the field samples as a function of stellar mass, using fit parameters from the combined ``full field'' sample tabulated in \autoref{table:mcmc_values}, and color-coding the galaxies as either quiescent or star-forming. Adopting a host stellar mass division at $\log(M_\star/M_\odot) = 10$, as typically used in SN~Ia cosmology \citep{2010MNRAS.406..782S,2013A&A...560A..66R,2019ApJ...881...19J,2020MNRAS.494.4426S}, we see that nearly all of the field quiescent hosts are in the higher-mass bin. The differences in the population parameters $\alpha$, $\beta$, and $M_B$ manifest themselves as a Hubble residual offset. We recover a mass step of $0.049 \pm 0.012$ mag, consistent with other low-redshift measurements \citep{2014A&A...568A..22B,2018ApJ...867..108J}, including recent investigations using SALT3 \citep{2023ApJ...951...22J}. Alternatively, we could derive a ``specific star-formation step'' of $0.049 \pm 0.011$ mag between the field quiescent and star-forming host SNe~Ia in \autoref{fig:mass_step}. If we allow for different $\alpha$, $\beta$, and $\sigma_\text{int}$ for these two samples and marginalize over them, the star-formation step is $0.057 \pm 0.013$ mag, given by the $M_B$ offset between the field quiescent and field star-forming galaxies in \autoref{table:mcmc_values}. It is reassuring that these slightly different approaches give a robust estimate of the step.

\autoref{fig:mass_step} also shows the weighted means of the Hubble residual of our \radius sample with the same host stellar mass division as the full field sample. We estimate the host masses of our \radius SNe Ia hosts using mass-to-light scaling relations of either their 2MASS $Ks$ or WISE $W1$ photometry \citep{2018ApJ...869...56B,2021ApJ...923..267D,2023ApJ...946...95J}, finding good consistency for objects with both 2MASS and WISE data. The lower mass bin \radius weighted mean Hubble residual is quite similar to the respective field sample weighted mean, while the higher mass bin \radius weighted average differs from the respective field sample mean by 0.045 $\pm$ 0.029 mag, i.e., consistent to within 2$\sigma$. One interesting trend to note is that our inner cluster quiescent hosts extend to lower mass than the field quiescent sample; we postulate that this may be due to stronger quenching of low-mass galaxies within clusters in the local Universe \citep{2010A&A...524A..76B,2018MNRAS.479.2147M}.

\citet{2023ApJ...953...35G} recently suggested that fast-declining SN~Ia may have a steeper dependence of luminosity on \xone, i.e., an $\alpha$ that varied with $\xone$. In \autoref{fig:alpha_fits} we examine this possibility with our inner cluster and field samples, plotting the SALT3 color-corrected SN~Ia absolute magnitudes (with $\beta = 2.395$ fixed from the field sample) versus \xone. We then perform a weighted linear regression for both faster- and slower-decliners separately (\xone $< -1$ and \xone $> -1$ respectively). We find $\alpha = -0.203 \pm 0.020$ for the faster-decliners and $\alpha = -0.087 \pm 0.014$ for the slower-decliners. These values are consistent to those from \citet{2023ApJ...953...35G} to within 1$\sigma$, confirming their finding favoring a nonlinear \xone correction rather than a single fixed $\alpha$. We also perform this same fit in three separate mass bins, shown in the right column panels of \autoref{fig:alpha_fits}. Here we do not confirm the suggestion of  \citet{2023ApJ...953...35G} that the steepening in $\alpha$ for \xone $< -1$ is stronger for higher stellar masses. Our $\alpha$ values for both the faster- and slower-declining samples are consistent to within 1$\sigma$ across the mass bins. Nevertheless, we concur with \citet{2023ApJ...953...35G} that a nonlinear correction in place of a single $\alpha$ value would benefit cosmological samples that contain SNe~Ia with a wide range of \xone.

\section{Discussion}\label{sec:discussion}

We have identified clear differences between samples of low-redshift SNe~Ia in cluster and field environments. It is important to interrogate to what extent our sample selection affects our results. For example, we have only included ``normal'' SNe~Ia, uniformly excluding any objects that were spectroscopically classified as 1991bg-like or 1991T/99aa-like from all samples. These would contribute to the fast-declining and slow-declining populations, respectively, and are further correlated with quiescent and star-forming environments. SN~Ia classifiers do not always provide this level of granularity; the quality and phase of the classification spectrum can affect whether a subtype designation can be determined, and there is a continuum between normal SN~Ia and these subtypes. Nevertheless, investigating the objects excluded by this selection, we find only one or two objects of each subclass for our cluster sample. Including them would not significantly alter our conclusions. 

Similarly, in our analysis we also uniformly excluded objects with $|x_1| > 3$, out of range of the SALT3 model. The number of objects rejected with $x_1 < -3$ or $x_1 > +3$ are 3/0, 0/0, 21/7, and 0/5, for our \radius, \tworadius, field quiescent, and field star-forming samples, respectively. If we assume the too-fast-declining objects are part of the fast-declining population (and conversely for the too-slow objects), our estimates of the fraction in the fast-declining population $f_1$ (see \autoref{fig:bimodal_fits}) would change by $+1.2\%$, $0.0\%$, $+3.0\%$, and $-0.2\%$ for those four samples. Standardization of fast-declining objects, especially relevant to the \radius sample may benefit from using other light curve fitting tools like MLCS2k2 or SNooPy \citep{2007ApJ...659..122J,Burns:2011,Burns:2014} that can better handle faster-evolving SNe~Ia.

The \xone distributions of the \radius and \tworadius samples are clearly bimodal (\autoref{fig:bimodal_fits}), and this bimodality is also strongly suggested in the field quiescent population. Recently, \citet{2023MNRAS.525.5187W} used a hierarchical Bayesian model to similarly identify two populations of SNe~Ia in the parameter space of $(x_1, c, m_B - \mu)$. As in our analysis, they find the greatest separation between the two populations in the \xone distributions and also note the correlation with host-galaxy properties. In their model the two populations furthermore have slightly different color ($c$) distributions, interpreted as arising from different intrinsic colors and dust reddening. We do not find conclusive evidence for differences in the $c$ distributions among our samples (\autoref{fig:param_comparison}), but this should be explored further, especially as our cluster (and even field quiescent) samples bring the two populations into much sharper relief compared to a full sample covering all environments.

Our field quiescent \xone distribution is consistent with other nearby samples of SNe~Ia in quiescent host galaxies \citep{2013A&A...560A..66R,2019JKAS...52..181K}, showing similar bimodality. Measurements at higher redshift tend to have more unimodal \xone distributions \citep{2013MNRAS.434.1443X,2022ApJ...938...62C}, typically with a higher mean \xone than we find here \citep[though perhaps excepting][]{2010ApJ...722..566L}. This could be a result of redshift evolution in quiescent galaxies and the SNe~Ia they host, but Malmquist bias may also be playing a role. 

Comparing our cluster SNe~Ia samples to those at higher redshift \citep{2013MNRAS.434.1443X,2023MNRAS.526.5292T}, we confirm the tendency of the cluster SNe~Ia to be faster evolving than their field counterparts. We also confirm with higher statistics the suggestions by \citet{2013MNRAS.434.1443X} that 1.~SNe~Ia closer to the cluster center have a higher fraction of fast-declining objects than farther out, 2.~cluster passive galaxies have a higher fraction of faster-declining objects than field passive galaxies, and 3.~the fast-declining cluster SNe~Ia are slightly more extreme (even lower \xone) than field quiescent SNe~Ia (\autoref{table:distribution_values}; \autoref{fig:bimodal_fits}). We caution that our study differs from the higher redshift examples because we are using an x-ray selected galaxy cluster catalog, while the high-redshift studies are limited to use optically-selected clusters. The potentially different cluster physical environments may play a role in the different SN~Ia populations we see and may also confound effects due to redshift evolution.

There are a few possibilities why SNe~Ia from \radius host galaxies may have different properties than SNe~Ia in the field, even restricting the samples to quiescent galaxies only. We note that whatever the cause, it must be intrinsic to the supernovae. Extrinsic factors like host-galaxy dust may affect the brightness or color of the SNe~Ia, but cannot alter the light curve shape ($x_1$) distribution to the large extent we see. 

Metallicity has been suggested as a factor in SN~Ia variation \citep{2011ApJ...743..172D,2013ApJ...770..108C}, and the deep gravitational potential well at the centers of galaxy clusters should retain metals better than field quiescent galaxies. However, studies of low-redshift cluster galaxies show they are only slightly more metal rich ($\lesssim$ 0.05 dex) than field counterparts \citep{Ellison:2009,Lara-Lopez:2022}. Observations of the intracluster medium do show a metal enhancement near low-$z$ cluster centers \citep{Lovisari:2019}, but the metals escape the galaxies (and subsequent generations of stars) similarly to field quiescent galaxies. It is unlikely, then, that differences in progenitor stellar metallicities are driving the differences seen in our cluster SNe~Ia.

The most likely explanation for the properties of the cluster SNe~Ia is the age of the stellar population from which they arise. Quiescent massive galaxies host fast-declining SNe~Ia, and these galaxies by definition will have preferentially older stars. A correlation between low $x_1$ and mean stellar age is expected and observed in field samples \citep{Gupta:2011,2020ApJ...889....8K} and age is implicated as the chief driver behind supernova standardization differences with host-galaxy properties \citep[recently in, e.g.,][]{2022A&A...657A..22B,Lee:2022,2023SCPMA..6629511W,2023MNRAS.520.6214W}. Turning to galaxy clusters, \cite{2013MNRAS.434.1443X} analyzed the ages of cluster SN Ia host galaxies and found that these hosts were older on average than field host galaxies. This is in accord with results that early type galaxies in nearby galaxy clusters are older than similar early type galaxies in the field by 1--2 Gyr \citep{2003ApJ...585...78V,2005ApJ...621..673T,2006ARA&A..44..141R}, an effect that has also been seen at higher redshifts \citep{2020MNRAS.498.5317W}. It is further intriguing that the strong shift to a fast-declining population in clusters is seen most clearly in our nearby sample, with a less pronounced trend at high redshift \citep{2023MNRAS.526.5292T}. This suggests that the fast-evolving population is tracing the oldest SN~Ia progenitors. 

The bimodality in the \xone distribution may be a hint that there is a qualitative difference between objects in the fast-evolving population and others. A more gradual evolution in the progenitor population might be more compatible with a gradual shift in a unimodal \xone distribution, but that is not what we observe. Given the wide range of possible SN~Ia progenitors and explosion mechanisms, it is enticing to speculate the \xone bimodality is a signal of the emergence of a different SN~Ia progenitor scenario in the oldest stellar populations. In a single-degenerate model with a Chandrasekhar-mass C/O white dwarf, for instance, the oldest SN~Ia have red giant companions \citep[see][for a review]{Maoz:2014}. Conversely, the delay times in typical double-degenerate SN~Ia models reflect the initial separation distribution for binary white dwarfs \citep{2018MNRAS.476.2584M}; if this is a power law as conventionally assumed, the qualitatively different behavior we observe for the oldest SNe~Ia might be unexpected.

Our results strengthen the case that older stellar populations produce atypical SNe~Ia \citep{2017ApJ...837..121G,2017ApJ...851L..50S,2020MNRAS.499.1424H,2021MNRAS.505L..52H,2022MNRAS.517L.132K,2023MNRAS.520L..21B}. The bimodality of the \xone distribution also suggests that caution is warranted in assuming SNe~Ia from old populations (in passive galaxies, for example) are continuously connected to those from younger populations. Though we do not find evidence for large differences in SN~Ia standardized luminosity that could depend on age (especially in our cluster objects; \autoref{fig:mass_step}), deriving age-dependent corrections from a passive galaxy sample (or potentially disregarding cluster versus field distinctions) may lead to results that are not applicable to the majority of SNe~Ia in star-forming galaxies or otherwise younger environments \citep{2019JKAS...52..181K,2019ApJ...874...32R,2020ApJ...896L...4R,2020ApJ...889....8K,2020ApJ...903...22L,Lee:2022,Murakami:2021,2023SCPMA..6629511W}.

\section{Summary and conclusions}\label{sec:summary}

Using archival data, we have constructed the largest to date sample of SNe Ia that occur within rich, nearby ($z < 0.1$), x-ray selected clusters of galaxies. We divide them into \radius (projected $r/r_{500} < 1$) and \tworadius ($1 < r/r_{500} < 2$) samples and compare these to samples of SNe~Ia in field quiescent and star-forming galaxies. With SALT3 light-curve fits to archival optical photometry, the cluster samples show a strongly bimodal distribution in light curve shape (SALT3 $x_1$) and we find a significant difference in the population of fast-evolving (low $x_1 < -1$) SNe~Ia in the clusters compared to field galaxies. Our \radius sample contains a much higher fraction of fast-evolving objects compared to the \tworadius sample or even a sample of field quiescent galaxies. These in turn have a higher fast-evolving fraction than field star-forming galaxies. We find no strong evidence of differences in the color (SALT3 $c$) distribution between the samples, and relatively small differences in standardization parameters ($\alpha$, $\beta$, $M_B$, $\sigma_\text{int}$) and standardized luminosities (Hubble residual). 

A key takeaway is that environmental correlations in SN~Ia properties extend beyond galactic scales: the \radius sample of SNe~Ia is intrinsically different from \tworadius objects and this difference persists even when comparing \radius quiescent host galaxies with \tworadius or field quiescent hosts. We suggest that the age of the stellar population is the more direct explanatory cause of these results, with the oldest stellar populations producing almost exclusively fast-evolving SNe~Ia. Future work can better clarify the galactic and local differences between low-redshift \radius SNe~Ia and other samples. Direct measurements and comparison of stellar ages (and perhaps metallicities) at the positions of these \radius supernovae and for their host galaxies should yield insight. 

We have shown that large samples of cluster SNe~Ia provide a unique window into SN~Ia populations and encourage further such observations, including extending to higher redshift. Upcoming large sky-area surveys, like the Vera C. Rubin Observatory Legacy Survey of Space and Time (LSST) will be transformative, allowing orders-of-magnitude increase in sample size and higher-significance determinations of sample differences. Moreover, a continued focus on nearby cluster SNe~Ia will be important, not only because these can be the best studied, but also they crucially come from the oldest populations. Such objects will likely be the key to unveiling the causal mechanism at work (e.g., different progenitor scenarios).  

While we do not see strong trends affecting Hubble residuals in our cluster samples, our results nevertheless have implications for SN~Ia cosmology. Our cluster SNe~Ia show somewhat higher scatter on the Hubble diagram than the field star-forming sample (\autoref{table:mcmc_values}), so excluding these (few in number) cluster objects would slightly improve cosmological samples. More worrisome are systematic uncertainties, especially if older stellar populations systematically produce different SNe~Ia. The oldest supernovae at any redshift are those with delay times approaching the age of the Universe at that redshift, a clearly evolving quantity. Conversely, the youngest SNe~Ia should have similar ages at all redshifts. Isolating these supernovae, by restricting samples to star-forming host galaxies, for instance, may prove a helpful strategy to reduce both statistical and systematic uncertainties for cosmology.

\section*{Acknowledgements}

We thank Andrew Baker, Yu-Yen Chang, Ryan Foley, and Jack Hughes for helpful discussions. We are grateful to Erik Peterson for help with the peculiar velocity maps used in our analysis.

C.L. acknowledges support from the National Science Foundation Graduate Research Fellowship under grant No. DGE-2233066.
S.W.J. is grateful for support of ground-based supernova cosmology research at Rutgers University through DOE award DE-SC0010008.
L.A.K. acknowledges support by NASA FINESST fellowship 80NSSC22K1599.

This research has made use of the NASA/IPAC Extragalactic Database (NED),
which is operated by the Jet Propulsion Laboratory, California Institute of Technology,
under contract with the National Aeronautics and Space Administration. 

The SALT spectra used in this study were obtained through the Rutgers University SALT program 2023-1-MLT-008 (PI: Jha).

The ZTF forced-photometry service was funded under the Heising-Simons Foundation grant
\#12540303 (PI: Graham).

\section*{Data Availability}

The data used in this study can be found at \url{https://github.com/Conor-Larison/cluster_sneia} and in Zenodo at doi:\href{https://zenodo.org/records/10150675}{10.5281/zenodo.10150675}.

\software{Jupyter \citep{Beg_2021}, Astropy \citep{2013A&A...558A..33A,2018AJ....156..123A}, Matplotlib \citep{2007CSE.....9...90H}, NumPy \citep{2020Natur.585..357H}, pandas \citep{mckinney-proc-scipy-2010,reback2020pandas}, SciPy \citep{2020NatMe..17..261V}, emcee \citep{2013PASP..125..306F}, corner \citep{corner}, \texttt{IRAF} \citep{1986SPIE..627..733T}, Pyraf \citep{2012SASS...31..159G}}

\bibliography{cluster_environ}{}
\bibliographystyle{aasjournal}

\appendix
\section{Appendix: Cluster Supernova Data}

\begin{longrotatetable}
\movetabledown=18mm
\tabletypesize{\footnotesize}

\begin{deluxetable}{lcccccccclccc}

\tablehead{
\colhead{SN} & \dcolhead{m_B \; \rm{(mag)}} & \dcolhead{x_1} & \dcolhead{c} & \dcolhead{\mu \; \rm{(mag)}} & \colhead{Cluster} & \colhead{$z$} & \dcolhead{r_{500} \; \rm{(Mpc)}} & \colhead{Proj.~$r/r_{500}$} & \colhead{NED Host Galaxy} & \colhead{Ks/W1 $log(M_{\star}/M_{\odot})$} & \colhead{Host/SN $z$} & \colhead{Host SFR}
}

\startdata
1990N    & $12.627 \pm 0.006$ & $+1.034 \pm 0.033$ & $+0.059 \pm 0.003$ & $31.953 \pm 0.784$ & MCXC J1230.7+1220 & 0.0036    & 0.75            & 1.11                  & NGC 4639                  & $8.906 \pm 0.046$                & 0.0034    & SF       \\
1992A    & $12.524 \pm 0.006$ & $-0.748 \pm 0.066$ & $+0.001 \pm 0.004$ & $31.670 \pm 0.422$ & MCXC J0338.4-3526 & 0.0051    & 0.40            & 0.60                  & NGC 1380                  & $10.154 \pm 0.046$               & 0.0063    & Q        \\
1994D    & $11.669 \pm 0.002$ & $-1.803 \pm 0.006$ & $-0.084 \pm 0.001$ & $30.994 \pm 0.784$ & MCXC J1230.7+1220 & 0.0036    & 0.75            & 1.68                  & NGC 4526                  & $9.870 \pm 0.006$                & 0.0021    & Q        \\
2000dk   & $15.245 \pm 0.003$ & $-2.379 \pm 0.027$ & $-0.034 \pm 0.002$ & $34.256 \pm 0.201$ & MCXC J0107.4+3227 & 0.0170    & 0.52            & 0.14                  & NGC 0382                  & $10.290 \pm 0.004$               & 0.0174    & Q        \\
2004fz   & $14.817 \pm 0.004$ & $-1.444 \pm 0.020$ & $+0.018 \pm 0.002$ & $33.931 \pm 0.209$ & MCXC J0200.2+3126 & 0.0168    & 0.46            & 1.27                  & NGC 0783                  & $10.148 \pm 0.044$               & 0.0173    & SF       \\
2005eu   & $16.319 \pm 0.006$ & $+0.482 \pm 0.050$ & $-0.138 \pm 0.004$ & $35.959 \pm 0.166$ & MCXC J0228.1+2811 & 0.0353    & 0.61            & 0.41                  & WISEA J022743.32+281037.7 & $9.394 \pm 0.058$                & 0.0345    & SF       \\
2006hx   & $17.402 \pm 0.004$ & $-0.299 \pm 0.024$ & $+0.114 \pm 0.003$ & $36.417 \pm 0.172$ & MCXC J0115.2+0019 & 0.0450    & 0.75            & 1.34                  & 2MASX J01135716+0022171   & $10.528 \pm 0.016$               & 0.0454    & Q        \\
2007ci   & $15.814 \pm 0.005$ & $-2.782 \pm 0.029$ & $+0.005 \pm 0.004$ & $34.678 \pm 0.183$ & MCXC J1144.6+1945 & 0.0214    & 0.90            & 0.47                  & NGC 3873                  & $10.545 \pm 0.042$               & 0.018     & Q        \\
2007fr   & $18.072 \pm 0.011$ & $-2.336 \pm 0.072$ & $-0.019 \pm 0.010$ & $37.052 \pm 0.161$ & MCXC J2137.1+0026 & 0.0509    & 0.41            & 0.10                  & \nodata                       & \nodata                              & 0.049     & \nodata      \\
2007nq   & $17.240 \pm 0.005$ & $-1.871 \pm 0.022$ & $-0.044 \pm 0.003$ & $36.454 \pm 0.171$ & MCXC J0056.3-0112 & 0.0442    & 0.94            & 1.19                  & UGC 00595                 & $10.913 \pm 0.044$               & 0.045     & Q        \\
2007on   & $12.958 \pm 0.001$ & $-2.065 \pm 0.005$ & $+0.004 \pm 0.001$ & $31.921 \pm 0.421$ & MCXC J0338.4-3526 & 0.0051    & 0.40            & 0.14                  & NGC 1404                  & $10.348 \pm 0.046$               & 0.0065    & Q        \\
2008L    & $15.149 \pm 0.008$ & $-1.825 \pm 0.033$ & $-0.117 \pm 0.005$ & $34.431 \pm 0.200$ & MCXC J0319.7+4130 & 0.0179    & 1.29            & 0.50                  & NGC 1259                  & $10.128 \pm 0.046$               & 0.0194    & Q        \\
2008Q    & $13.367 \pm 0.005$ & $-1.761 \pm 0.069$ & $-0.032 \pm 0.004$ & $32.457 \pm 0.355$ & MCXC J0124.8+0932 & 0.0079    & 0.21            & 0.11                  & NGC 0524                  & $10.221 \pm 0.046$               & 0.008     & Q        \\
2008bf   & $15.631 \pm 0.005$ & $+0.395 \pm 0.038$ & $-0.110 \pm 0.003$ & $35.192 \pm 0.176$ & MCXC J1204.1+2020 & 0.0252    & 0.48            & 0.41                  & NGC 4055                  & $10.821 \pm 0.044$               & 0.0244    & Q        \\
2009eu   & $17.648 \pm 0.032$ & $-2.360 \pm 0.190$ & $+0.237 \pm 0.021$ & $36.016 \pm 0.174$ & MCXC J1628.6+3932 & 0.0299    & 1.00            & 0.02                  & LEDA 3084828              & $10.116 \pm 0.036$               & 0.0292    & Q        \\
2010ai   & $15.920 \pm 0.010$ & $-1.685 \pm 0.048$ & $-0.070 \pm 0.007$ & $35.110 \pm 0.183$ & MCXC J1259.7+2756 & 0.0231    & 1.14            & 0.13                  & WISEA J125925.01+275948.2 & $8.859 \pm 0.066$                & 0.0183    & Q        \\
2012cg   & $11.982 \pm 0.002$ & $+0.526 \pm 0.010$ & $+0.125 \pm 0.001$ & $31.078 \pm 0.784$ & MCXC J1230.7+1220 & 0.0036    & 0.75            & 1.09                  & NGC 4424                  & $8.619 \pm 0.046$                & 0.0015    & SF       \\
2013cs   & $13.671 \pm 0.001$ & $+0.768 \pm 0.010$ & $+0.038 \pm 0.001$ & $33.016 \pm 0.287$ & MCXC J1315.3-1623 & 0.0087    & 0.56            & 1.80                  & ESO 576- G 017            & $9.315 \pm 0.058$                & 0.0092    & Q        \\
2014ai   & $16.258 \pm 0.035$ & $-1.447 \pm 0.184$ & $+0.089 \pm 0.021$ & $35.100 \pm 0.188$ & MCXC J0919.8+3345 & 0.0230    & 0.45            & 0.07                  & NGC 2832                  & $10.885 \pm 0.042$               & 0.023     & Q        \\
2015ar   & $14.932 \pm 0.023$ & $-2.055 \pm 0.153$ & $-0.080 \pm 0.029$ & $34.097 \pm 0.210$ & MCXC J0107.4+3227 & 0.0170    & 0.52            & 0.16                  & NGC 383                   & $10.748 \pm 0.044$               & 0.017     & Q        \\
2018bgs  & $18.253 \pm 0.005$ & $-0.075 \pm 0.070$ & $-0.086 \pm 0.005$ & $37.696 \pm 0.158$ & MCXC J1421.5+4933 & 0.0716    & 0.82            & 0.03                  & MCG +08-26-021            & $10.808 \pm 0.048$               & 0.0719    & Q        \\
2018ccl  & $16.772 \pm 0.002$ & $-1.582 \pm 0.015$ & $+0.094 \pm 0.002$ & $35.584 \pm 0.170$ & MCXC J1628.6+3932 & 0.0299    & 1.00            & 0.61                  & UGC 10404                 & $10.324 \pm 0.046$               & 0.0268    & SF       \\
2018cng  & $18.341 \pm 0.007$ & $+0.794 \pm 0.068$ & $+0.141 \pm 0.006$ & $37.429 \pm 0.168$ & MCXC J1545.0+3603 & 0.0654    & 0.80            & 1.31                  & WISEA J154529.02+355118.7 & $9.274 \pm 0.084$                & 0.0661    & Q        \\
2018dvb  & $17.903 \pm 0.006$ & $-1.541 \pm 0.045$ & $-0.113 \pm 0.006$ & $37.213 \pm 0.158$ & MCXC J1533.2+3108 & 0.0673    & 0.67            & 0.99                  & WISEA J153356.74+311009.7 & $10.493 \pm 0.080$               & 0.0649    & Q        \\
2018eak  & $16.908 \pm 0.006$ & $-2.660 \pm 0.052$ & $+0.030 \pm 0.005$ & $35.730 \pm 0.171$ & MCXC J1615.5+1927 & 0.0308    & 0.50            & 0.91                  & 2MASX J16154858+1939440   & $9.844 \pm 0.044$                & 0.0303    & Q        \\
2018fio  & $18.992 \pm 0.039$ & $-1.758 \pm 0.332$ & $+0.189 \pm 0.029$ & $37.554 \pm 0.163$ & MCXC J1521.2+3038 & 0.0777    & 0.98            & 0.05                  & WISEA J152113.56+303813.2 & $9.599 \pm 0.072$                & 0.075     & Q        \\
2018ggt  & $17.000 \pm 0.009$ & $-2.368 \pm 0.052$ & $-0.078 \pm 0.007$ & $36.236 \pm 0.172$ & MCXC J0056.3-0112 & 0.0442    & 0.94            & 1.30                  & UGC 00588                 & $11.136 \pm 0.044$               & 0.0442    & Q        \\
2018hts  & $16.562 \pm 0.015$ & $+0.531 \pm 0.125$ & $-0.095 \pm 0.013$ & $36.212 \pm 0.174$ & MCXC J2323.8+1648 & 0.0416    & 0.88            & 1.10                  & WISEA J232432.94+170539.3 & $8.496 \pm 0.247$                & 0.04      & SF       \\
2018hzd  & $17.891 \pm 0.008$ & $+0.130 \pm 0.125$ & $+0.018 \pm 0.007$ & $37.204 \pm 0.170$ & MCXC J2336.5+2108 & 0.0565    & 0.84            & 1.46                  & WISEA J233701.22+212551.7 & $8.951 \pm 0.156$                & 0.05      & Q        \\
2018jsv  & $18.817 \pm 0.019$ & $-2.986 \pm 0.188$ & $-0.062 \pm 0.017$ & $37.816 \pm 0.162$ & MCXC J1111.6+4050 & 0.0794    & 0.87            & 0.19                  & MCG +07-23-031            & $11.576 \pm 0.013$               & 0.0781    & Q        \\
2019aex  & $17.011 \pm 0.025$ & $-2.279 \pm 0.382$ & $-0.045 \pm 0.027$ & $36.063 \pm 0.173$ & MCXC J0040.0+0649 & 0.0395    & 0.70            & 0.91                  & WISEA J003915.70+064118.6 & $10.647 \pm 0.044$               & 0.0387    & Q        \\
2019bsa  & $16.871 \pm 0.010$ & $-2.277 \pm 0.047$ & $-0.025 \pm 0.006$ & $35.874 \pm 0.171$ & MCXC J1110.7+2842 & 0.0314    & 0.59            & 0.07                  & MCG +05-27-004 NED02      & $10.295 \pm 0.044$               & 0.0339    & Q        \\
2019bxi  & $18.564 \pm 0.025$ & $-2.539 \pm 0.159$ & $+0.132 \pm 0.018$ & $37.248 \pm 0.172$ & MCXC J1107.3-2300 & 0.0639    & 0.73            & 1.99                  & WISEA J110628.61-224427.3 & $10.518 \pm 0.058$               & 0.0637    & SF       \\
2019cdn  & $17.566 \pm 0.005$ & $-1.471 \pm 0.045$ & $+0.071 \pm 0.004$ & $36.542 \pm 0.171$ & MCXC J1010.2+5429 & 0.0470    & 0.51            & 1.85                  & CGCG 266-030              & $10.827 \pm 0.044$               & 0.0464    & Q        \\
2019cmx  & $18.551 \pm 0.007$ & $+0.600 \pm 0.096$ & $-0.021 \pm 0.007$ & $37.928 \pm 0.157$ & MCXC J1654.7+5854 & 0.0869    & 0.70            & 0.31                  & WISEA J165500.60+585522.3 & $9.487 \pm 0.080$                & 0.092     & SF       \\
2019cnu  & $17.194 \pm 0.004$ & $-2.759 \pm 0.028$ & $+0.223 \pm 0.004$ & $35.542 \pm 0.175$ & MCXC J1755.8+6236 & 0.0266    & 0.47            & 0.01                  & NGC 6521                  & $11.333 \pm 0.005$               & 0.0271    & Q        \\
2019dom  & $18.595 \pm 0.036$ & $-1.175 \pm 0.196$ & $+0.150 \pm 0.025$ & $37.328 \pm 0.160$ & MCXC J1834.1+7057 & 0.0824    & 0.70            & 0.03                  & WISEA J183408.56+705719.3 & $11.168 \pm 0.046$               & 0.0834    & Q        \\
2019dye  & $17.422 \pm 0.022$ & $-0.620 \pm 0.339$ & $-0.007 \pm 0.018$ & $36.704 \pm 0.178$ & MCXC J0602.0+5315 & 0.0510    & 0.72            & 1.29                  & WISEA J060024.20+532132.6 & $9.235 \pm 0.090$                & 0.047     & Q        \\
2019fcp  & $17.585 \pm 0.043$ & $-2.037 \pm 0.469$ & $+0.094 \pm 0.027$ & $36.338 \pm 0.175$ & MCXC J0737.6+5920 & 0.0384    & 0.47            & 0.16                  & UGC 03928                 & $10.784 \pm 0.044$               & 0.0387    & Q        \\
2019gtl  & $18.288 \pm 0.028$ & $+0.441 \pm 0.389$ & $-0.021 \pm 0.022$ & $37.644 \pm 0.168$ & MCXC J1113.3+0231 & 0.0780    & 0.87            & 0.79                  & WISEA J111315.84+022415.5 & $9.249 \pm 0.177$                & 0.079     & Q        \\
2019gwn  & $18.216 \pm 0.057$ & $+0.430 \pm 0.865$ & $-0.087 \pm 0.045$ & $37.726 \pm 0.208$ & MCXC J2058.2-0745 & 0.0793    & 0.75            & 0.99                  & SALT TARGET               & \nodata                              & 0.0816    & SF       \\
2019hhz  & $18.405 \pm 0.030$ & $-2.465 \pm 0.150$ & $+0.070 \pm 0.026$ & $37.159 \pm 0.167$ & MCXC J2350.8+0609 & 0.0562    & 0.86            & 0.34                  & WISEA J235038.57+060626.0 & $9.939 \pm 0.052$                & 0.0553    & Q        \\
2019krv  & $16.153 \pm 0.005$ & $-2.304 \pm 0.020$ & $+0.018 \pm 0.003$ & $35.051 \pm 0.179$ & MCXC J2214.8+1350 & 0.0253    & 0.48            & 0.15                  & NGC 7236                  & $10.548 \pm 0.046$               & 0.0262    & Q        \\
2019mbs  & $17.090 \pm 0.022$ & $-2.145 \pm 0.134$ & $+0.025 \pm 0.019$ & $35.993 \pm 0.170$ & MCXC J0338.6+0958 & 0.0347    & 1.05            & 0.36                  & WISEA J033814.09+100503.5 & $11.204 \pm 0.203$               & 0.0382    & Q        \\
2019qff  & $18.314 \pm 0.028$ & $-0.321 \pm 0.244$ & $+0.062 \pm 0.023$ & $37.371 \pm 0.164$ & MCXC J0015.4-2350 & 0.0645    & 0.69            & 0.97                  & WISEA J001549.99-235727.7 & $10.230 \pm 0.052$               & 0.063     & SF       \\
2019rzm  & $16.828 \pm 0.016$ & $-0.519 \pm 0.155$ & $-0.075 \pm 0.012$ & $36.295 \pm 0.174$ & MCXC J0116.1-1555 & 0.0448    & 0.55            & 1.81                  & MCG -03-04-038            & $10.822 \pm 0.046$               & 0.0448    & Q        \\
2019sen  & $16.934 \pm 0.007$ & $+0.343 \pm 0.058$ & $-0.052 \pm 0.006$ & $36.450 \pm 0.170$ & MCXC J1811.0+4954 & 0.0501    & 0.77            & 1.35                  & WISEA J181016.57+501048.2 & $9.886 \pm 0.050$                & 0.046     & Q        \\
2019ulw  & $17.277 \pm 0.005$ & $+0.103 \pm 0.054$ & $-0.058 \pm 0.005$ & $36.780 \pm 0.171$ & MCXC J0246.0+3653 & 0.0488    & 0.81            & 1.02                  & \nodata                       & \nodata                              & 0.044     & \nodata      \\
2019uyw  & $18.401 \pm 0.042$ & $-0.568 \pm 0.375$ & $-0.113 \pm 0.032$ & $37.956 \pm 0.182$ & MCXC J0257.8+1302 & 0.0722    & 1.12            & 1.13                  & WISEA J025654.38+131012.3 & $10.300 \pm 0.050$               & 0.081     & Q        \\
2019wmn  & $18.343 \pm 0.024$ & $-1.227 \pm 0.176$ & $-0.047 \pm 0.019$ & $37.538 \pm 0.162$ & MCXC J1039.4+0510 & 0.0700    & 0.81            & 0.43                  & WISEA J103911.23+050945.6 & $9.857 \pm 0.066$                & 0.0708    & Q        \\
2020ags  & $15.543 \pm 0.001$ & $+0.319 \pm 0.008$ & $-0.062 \pm 0.001$ & $35.083 \pm 0.191$ & MCXC J1259.7+2756 & 0.0231    & 1.14            & 1.90                  & \nodata                       & \nodata                              & 0.02      & \nodata      \\
2020cox  & $18.861 \pm 0.018$ & $-2.129 \pm 0.187$ & $+0.068 \pm 0.016$ & $37.662 \pm 0.161$ & MCXC J1217.6+0339 & 0.0766    & 1.05            & 0.65                  & WISEA J121709.82+033806.7 & $10.653 \pm 0.074$               & 0.0788    & Q        \\
2020ddo  & $17.261 \pm 0.014$ & $-1.503 \pm 0.114$ & $-0.023 \pm 0.011$ & $36.470 \pm 0.174$ & MCXC J1539.6+2147 & 0.0411    & 0.80            & 1.78                  & WISEA J154107.78+220734.0 & $10.093 \pm 0.046$               & 0.0422    & Q        \\
2020ftd  & $17.659 \pm 0.046$ & $-1.749 \pm 0.163$ & $-0.004 \pm 0.029$ & $36.683 \pm 0.165$ & MCXC J1253.2-1522 & 0.0462    & 0.69            & 0.39                  & WISEA J125303.79-152701.4 & $10.530 \pm 0.044$               & 0.043     & Q        \\
2020iyz  & $17.601 \pm 0.013$ & $-2.431 \pm 0.101$ & $+0.010 \pm 0.011$ & $36.501 \pm 0.163$ & MCXC J1257.1-1724 & 0.0473    & 0.99            & 0.73                  & WISEA J125751.10-171537.4 & $9.824 \pm 0.054$                & 0.05      & Q        \\
2020jee  & $15.402 \pm 0.003$ & $-1.552 \pm 0.019$ & $-0.119 \pm 0.002$ & $34.847 \pm 0.192$ & MCXC J1144.6+1945 & 0.0214    & 0.90            & 1.12                  & CGCG 097-064              & $9.565 \pm 0.048$                & 0.0198    & SF       \\
2020jny  & $16.184 \pm 0.003$ & $-1.655 \pm 0.019$ & $-0.060 \pm 0.003$ & $35.466 \pm 0.184$ & MCXC J1204.1+2020 & 0.0252    & 0.48            & 1.48                  & WISEA J120258.54+200506.9 & \nodata                              & 0.0237    & GV       \\
2020kpw  & $17.199 \pm 0.020$ & $+0.813 \pm 0.156$ & $-0.033 \pm 0.014$ & $36.728 \pm 0.171$ & MCXC J2158.3-2006 & 0.0570    & 0.67            & 1.08                  & WISEA J215746.71-195834.3 & $9.923 \pm 0.058$                & 0.0568    & Q        \\
2020lxu  & $16.821 \pm 0.016$ & $-1.491 \pm 0.039$ & $+0.017 \pm 0.010$ & $35.829 \pm 0.167$ & MCXC J0036.5+2544 & 0.0341    & 0.55            & 0.45                  & WISEA J003608.66+254504.3 & $10.089 \pm 0.060$               & 0.0332    & Q        \\
2020nin  & $18.318 \pm 0.013$ & $-0.672 \pm 0.101$ & $+0.043 \pm 0.010$ & $37.468 \pm 0.169$ & MCXC J1825.3+3026 & 0.0650    & 0.95            & 1.28                  & WISEA J182606.42+301320.6 & $9.557 \pm 0.072$                & 0.06      & SF       \\
2020nlh  & $18.892 \pm 0.011$ & $-1.470 \pm 0.066$ & $-0.008 \pm 0.008$ & $37.962 \pm 0.156$ & MCXC J1620.5+2953 & 0.0972    & 0.94            & 0.88                  & WISEA J162030.13+300121.4 & $10.422 \pm 0.052$               & 0.0953    & Q        \\
2020ppe  & $17.733 \pm 0.009$ & $-2.918 \pm 0.048$ & $+0.143 \pm 0.006$ & $36.251 \pm 0.164$ & MCXC J0040.0+0649 & 0.0395    & 0.70            & 0.75                  & WISEA J003936.06+063951.2 & \nodata                              & 0.0389    & Q        \\
2020sia  & $18.894 \pm 0.015$ & $+0.401 \pm 0.105$ & $-0.024 \pm 0.010$ & $38.350 \pm 0.167$ & MCXC J0020.6+2840 & 0.0940    & 0.92            & 1.01                  & WISEA J002040.60+284835.4 & $9.151 \pm 0.296$                & 0.094     & Q        \\
2020vnr  & $18.514 \pm 0.013$ & $+0.014 \pm 0.154$ & $-0.076 \pm 0.011$ & $38.050 \pm 0.169$ & MCXC J0003.8+0203 & 0.0924    & 0.82            & 1.78                  & \nodata                       & \nodata                              & 0.09      & \nodata      \\
2020wcj  & $15.959 \pm 0.003$ & $-1.649 \pm 0.012$ & $+0.085 \pm 0.002$ & $34.784 \pm 0.182$ & MCXC J0252.8-0116 & 0.0235    & 0.54            & 0.14                  & \nodata                       & \nodata                              & 0.0237    & \nodata      \\
2020xps  & $18.058 \pm 0.017$ & $+0.272 \pm 0.121$ & $-0.053 \pm 0.012$ & $37.570 \pm 0.169$ & MCXC J0258.9+1334 & 0.0739    & 1.24            & 1.44                  & WISEA J025733.72+132831.6 & $10.109 \pm 0.056$               & 0.07      & SF       \\
2020yji  & $17.711 \pm 0.013$ & $+1.428 \pm 0.124$ & $-0.072 \pm 0.010$ & $37.320 \pm 0.160$ & MCXC J1200.3+5613 & 0.0650    & 0.76            & 0.96                  & \nodata                       & \nodata                              & 0.07      & \nodata      \\
2020zgh  & $17.324 \pm 0.013$ & $-0.327 \pm 0.087$ & $+0.082 \pm 0.010$ & $36.418 \pm 0.172$ & MCXC J0115.2+0019 & 0.0450    & 0.75            & 1.52                  & WISEA J011553.37+000056.8 & $9.115 \pm 0.094$                & 0.045     & Q        \\
2020acqt & $18.268 \pm 0.012$ & $+0.792 \pm 0.085$ & $+0.169 \pm 0.010$ & $37.287 \pm 0.171$ & MCXC J1147.3+5544 & 0.0510    & 0.62            & 2.00                  & MRK 1455                  & $10.461 \pm 0.046$               & 0.0529    & SF       \\
2020acwj & $17.421 \pm 0.023$ & $-0.399 \pm 0.134$ & $+0.224 \pm 0.016$ & $36.147 \pm 0.173$ & MCXC J2310.4+0734 & 0.0424    & 0.73            & 1.48                  & WISEA J231138.74+074649.1 & $9.179 \pm 0.082$                & 0.04      & Q        \\
2020aden & $17.878 \pm 0.029$ & $-2.021 \pm 0.144$ & $-0.075 \pm 0.021$ & $37.034 \pm 0.162$ & MCXC J0108.8-1524 & 0.0533    & 0.75            & 0.98                  & WISEA J010801.10-152419.6 & $9.724 \pm 0.060$                & 0.055     & Q        \\
2021wh   & $17.024 \pm 0.010$ & $-2.098 \pm 0.058$ & $+0.032 \pm 0.008$ & $36.018 \pm 0.177$ & MCXC J0937.9-2020 & 0.0344    & 0.55            & 1.06                  & WISEA J093754.93-200640.9 & $8.652 \pm 0.110$                & 0.035     & SF       \\
2021xy   & $17.709 \pm 0.039$ & $-2.463 \pm 0.372$ & $-0.031 \pm 0.032$ & $36.702 \pm 0.169$ & MCXC J0102.7-2152 & 0.0569    & 0.94            & 0.20                  & WISEA J010249.23-215010.6 & $9.933 \pm 0.058$                & 0.0609    & Q        \\
2021ajy  & $18.186 \pm 0.029$ & $-1.807 \pm 0.130$ & $-0.192 \pm 0.020$ & $37.650 \pm 0.160$ & MCXC J1115.5+5426 & 0.0691    & 0.70            & 0.55                  & WISEA J111605.64+542713.4 & $10.141 \pm 0.052$               & 0.0675    & Q        \\
2021aut  & $17.044 \pm 0.018$ & $+0.755 \pm 0.123$ & $-0.019 \pm 0.011$ & $36.533 \pm 0.172$ & MCXC J1326.9-2710 & 0.0458    & 0.97            & 1.39                  & LEDA 764003               & $9.587 \pm 0.056$                & 0.047     & SF       \\
2021bbz  & $15.555 \pm 0.003$ & $-0.149 \pm 0.030$ & $+0.019 \pm 0.002$ & $34.831 \pm 0.192$ & MCXC J1144.6+1945 & 0.0214    & 0.90            & 1.03                  & ARK 321                   & $10.448 \pm 0.046$               & 0.0233    & Q        \\
2021bjd  & $17.686 \pm 0.009$ & $+0.825 \pm 0.113$ & $-0.115 \pm 0.006$ & $37.424 \pm 0.169$ & MCXC J1348.8+2635 & 0.0622    & 1.22            & 1.09                  & SDSS J134749.22+262400.4  & \nodata                              & 0.068     & Q        \\
2021bsf  & $16.158 \pm 0.003$ & $-1.780 \pm 0.023$ & $-0.044 \pm 0.002$ & $35.384 \pm 0.184$ & MCXC J1204.1+2020 & 0.0252    & 0.48            & 1.83                  & WISEA J120604.02+203213.0 & $8.959 \pm 0.068$                & 0.0237    & Q        \\
2021cai  & $16.731 \pm 0.031$ & $-2.684 \pm 0.153$ & $-0.071 \pm 0.017$ & $35.790 \pm 0.169$ & MCXC J0036.5+2544 & 0.0341    & 0.55            & 1.00                  & CGCG 479-042              & $10.640 \pm 0.044$               & 0.0328    & Q        \\
2021fom  & $18.808 \pm 0.016$ & $-1.528 \pm 0.132$ & $-0.035 \pm 0.015$ & $37.935 \pm 0.161$ & MCXC J1259.3-0411 & 0.0845    & 1.13            & 0.20                  & WISEA J125912.06-041214.0 & $9.217 \pm 0.191$                & 0.09      & Q        \\
2021hiz  & $13.097 \pm 0.001$ & $-0.485 \pm 0.003$ & $+0.082 \pm 0.000$ & $32.170 \pm 0.784$ & MCXC J1230.7+1220 & 0.0036    & 0.75            & 1.88                  & IC 3322A                  & $9.296 \pm 0.010$                & 0.0033    & Q        \\
2021kaq  & $17.358 \pm 0.026$ & $-2.045 \pm 0.156$ & $+0.019 \pm 0.020$ & $36.392 \pm 0.176$ & MCXC J0748.1+1832 & 0.0400    & 0.56            & 1.54                  & WISEA J074920.72+183920.9 & $10.045 \pm 0.048$               & 0.0434    & Q        \\
2021kqo  & $18.760 \pm 0.015$ & $-1.752 \pm 0.159$ & $+0.234 \pm 0.013$ & $37.288 \pm 0.170$ & MCXC J1533.2+3108 & 0.0673    & 0.67            & 1.67                  & 2MASS J15335013+3056137   & $10.753 \pm 0.018$               & 0.0658    & Q        \\
2021low  & $15.452 \pm 0.002$ & $+0.811 \pm 0.013$ & $-0.040 \pm 0.002$ & $34.903 \pm 0.183$ & MCXC J1144.6+1945 & 0.0214    & 0.90            & 0.64                  & UGC 06719                 & $10.163 \pm 0.044$               & 0.0219    & SF       \\
2021qxq  & $18.876 \pm 0.014$ & $+0.923 \pm 0.136$ & $+0.032 \pm 0.012$ & $38.169 \pm 0.160$ & MCXC J1558.3+2713 & 0.0894    & 1.38            & 0.65                  & SDSS J155819.79+272234.0  & \nodata                              & 0.0886    & Q        \\
2021qyf  & $19.016 \pm 0.022$ & $-2.596 \pm 0.285$ & $+0.090 \pm 0.018$ & $37.801 \pm 0.174$ & MCXC J1342.0+0213 & 0.0765    & 0.82            & 1.93                  & WISEA J134122.10+015905.5 & $9.960 \pm 0.062$                & 0.0801    & SF       \\
2021wyw  & $17.250 \pm 0.005$ & $-2.023 \pm 0.059$ & $-0.030 \pm 0.005$ & $36.299 \pm 0.163$ & MCXC J2310.4+0734 & 0.0424    & 0.73            & 0.07                  & NGC 7501                  & $10.984 \pm 0.044$               & 0.0427    & Q        \\
2021ypa  & $16.742 \pm 0.012$ & $-2.048 \pm 0.051$ & $-0.006 \pm 0.009$ & $35.839 \pm 0.179$ & MCXC J0433.6-1315 & 0.0326    & 1.00            & 1.14                  & WISEA J043538.20-131323.3 & $9.742 \pm 0.052$                & 0.0344    & SF       \\
2021zfo  & $18.843 \pm 0.022$ & $-0.163 \pm 0.228$ & $+0.053 \pm 0.020$ & $37.941 \pm 0.164$ & MCXC J1800.5+6913 & 0.0823    & 0.69            & 0.50                  & WISEA J180027.09+690944.9 & $10.067 \pm 0.052$               & 0.085     & SF       \\
2021zgk  & $18.232 \pm 0.017$ & $-2.145 \pm 0.163$ & $+0.095 \pm 0.015$ & $36.968 \pm 0.162$ & MCXC J1857.6+3800 & 0.0567    & 0.84            & 0.36                  & WISEA J185715.87+375914.6 & $10.965 \pm 0.044$               & 0.0523    & Q        \\
2021aalx & $16.948 \pm 0.008$ & $-1.983 \pm 0.044$ & $+0.142 \pm 0.006$ & $35.593 \pm 0.167$ & MCXC J0036.5+2544 & 0.0341    & 0.55            & 0.59                  & UGC 00367 & $10.446 \pm 0.046$               & 0.0322    & SF       \\
2021abzf & $15.992 \pm 0.029$ & $-1.947 \pm 0.092$ & $+0.047 \pm 0.020$ & $34.968 \pm 0.192$ & MCXC J1144.6+1945 & 0.0214    & 0.90            & 1.75                  & WISEA J114854.88+194833.5 & $9.523 \pm 0.048$                & 0.0237    & Q        \\
2021accx & $16.080 \pm 0.004$ & $-0.123 \pm 0.051$ & $-0.067 \pm 0.003$ & $35.577 \pm 0.179$ & MCXC J2338.4+2700 & 0.0309    & 0.75            & 1.32                  & WISEA J234001.38+271627.4 & $8.227 \pm 0.221$                & 0.031     & Q        \\
2021achd & $17.992 \pm 0.010$ & $-2.028 \pm 0.145$ & $+0.037 \pm 0.009$ & $36.983 \pm 0.170$ & MCXC J0039.6+2114 & 0.0619    & 0.63            & 1.38                  & WISEA J003848.18+210933.6 & $10.322 \pm 0.050$               & 0.0601    & Q        \\
2022bij  & $17.954 \pm 0.027$ & $+1.764 \pm 0.194$ & $+0.168 \pm 0.016$ & $37.100 \pm 0.172$ & MCXC J0828.6+3025 & 0.0503    & 0.73            & 1.62                  & WISEA J082947.12+303921.4 & $9.682 \pm 0.060$                & 0.0489    & SF       \\
2022czi  & $18.765 \pm 0.029$ & $-2.041 \pm 0.271$ & $+0.047 \pm 0.025$ & $37.628 \pm 0.168$ & MCXC J1200.3+5613 & 0.0650    & 0.76            & 0.17                  & MCG +09-20-056            & $10.979 \pm 0.048$               & 0.0648    & Q        \\
2022mww  & $15.453 \pm 0.005$ & $-1.892 \pm 0.018$ & $+0.099 \pm 0.003$ & $34.305 \pm 0.208$ & MCXC J0319.7+4130 & 0.0179    & 1.29            & 1.37                  & WISEA J031247.23+414914.6 & $9.230 \pm 0.251$                & 0.02      & Q        \\
2022nzb  & $18.484 \pm 0.017$ & $-2.554 \pm 0.433$ & $+0.006 \pm 0.015$ & $37.378 \pm 0.173$ & MCXC J1351.7+4622 & 0.0625    & 0.67            & 0.99                  & CGCG 246-024              & $11.327 \pm 0.046$               & 0.0623    & Q        \\
2022rdt  & $18.811 \pm 0.022$ & $+0.192 \pm 0.207$ & $-0.060 \pm 0.017$ & $38.330 \pm 0.170$ & MCXC J0020.6+2840 & 0.0940    & 0.92            & 1.75                  & \nodata                       & \nodata                              & 0.094     & \nodata      \\
2022rjj  & $18.496 \pm 0.032$ & $-2.418 \pm 0.185$ & $+0.146 \pm 0.021$ & $37.162 \pm 0.172$ & MCXC J0041.8-0918 & 0.0555    & 1.21            & 1.49                  & WISEA J004028.94-093729.9 & $9.865 \pm 0.058$                & 0.0562    & Q        \\
2022rqc  & $16.464 \pm 0.013$ & $+1.065 \pm 0.051$ & $+0.125 \pm 0.008$ & $35.628 \pm 0.177$ & MCXC J0740.9+5525 & 0.0340    & 0.76            & 1.95                  & UGC 03923                 & $10.706 \pm 0.012$               & 0.03      & SF       \\
2022vrn  & $17.655 \pm 0.010$ & $+0.174 \pm 0.096$ & $-0.111 \pm 0.009$ & $37.189 \pm 0.160$ & MCXC J0753.4+2921 & 0.0621    & 0.78            & 0.97                  & SDSS J075249.11+292909.2  & \nodata                              & 0.06      & Q
\enddata
\tablecomments{In the Host SFR column, SF=star-forming, Q=quiescent, and GV=green valley.}
\end{deluxetable}

\end{longrotatetable}

\end{document}